\documentclass{aa}  

\usepackage[usenames, dvipsnames]{color}
\usepackage{graphicx}
\usepackage{txfonts}
\usepackage{amsmath}
\usepackage{textgreek}
\usepackage{natbib,twoopt}
\usepackage[colorlinks=true,allcolors=blue]{hyperref} 
\bibpunct{(}{)}{;}{a}{}{,} 
\makeatletter
\newcommandtwoopt{\citeads}[3][][]{\href{http://adsabs.harvard.edu/abs/#3}%
{\def\hyper@linkstart##1##2{}%
\let\hyper@linkend\@empty\citealp[#1][#2]{#3}}}
\newcommandtwoopt{\citepads}[3][][]{\href{http://adsabs.harvard.edu/abs/#3}%
{\def\hyper@linkstart##1##2{}%
\let\hyper@linkend\@empty\citep[#1][#2]{#3}}}
\newcommandtwoopt{\citetads}[3][][]{\href{http://adsabs.harvard.edu/abs/#3}%
{\def\hyper@linkstart##1##2{}%
\let\hyper@linkend\@empty\citet[#1][#2]{#3}}}
\newcommandtwoopt{\citeyearads}[3][][]%
{\href{http://adsabs.harvard.edu/abs/#3}
{\def\hyper@linkstart##1##2{}%
\let\hyper@linkend\@empty\citeyear[#1][#2]{#3}}}
\makeatother
\usepackage{siunitx}
\DeclareSIUnit \parsec {pc}
\DeclareSIUnit\angstrom{\text {Å}}
\DeclareSIUnit\year{\text {yr}}
\DeclareSIUnit\erg{\text {erg}}
\DeclareSIUnit \h {\ensuremath{\mathit{h}}}

\usepackage{ulem}

\usepackage{xcolor}

\usepackage{comment}

\newcommand{\Msun}{M_\odot}
\newcommand{\solarmass}{M_\odot}

\begin{document} 

\title{Black hole mergers as tracers of spinning massive black hole and galaxy populations in the \textsc{Obelisk} simulation}
   \titlerunning{Massive black hole mergers in the \textsc{Obelisk} simulation}


   \author{Chi An Dong-P{\'a}ez\inst{1}
          \and
          Marta Volonteri\inst{1}
          \and 
          Ricarda S. Beckmann\inst{2}
          \and 
          Yohan Dubois\inst{1}
          \and
          Maxime Trebitsch\inst{3}
          \and
          Alberto Mangiagli\inst{4}
          \and
          Susanna D. Vergani\inst{5,1}
          \and
          Natalie A. Webb\inst{6}
          }

   \institute{Institut d’Astrophysique de Paris, UMR 7095, 
                CNRS and Sorbonne Universit\'{e}, 98 bis boulevard Arago, 75014 Paris, France\\
              \email{dongpaez@iap.fr}
         \and
              Institute of Astronomy and Kavli Institute for Cosmology, University of Cambridge, Madingley Road, Cambridge, CB3 0HA, UK
         \and
             Kapteyn Astronomical Institute, University of Groningen, P.O. Box 800, 9700 AV Groningen, The Netherlands
         \and
             AstroParticule et Cosmologie, Universit{\'e} Paris, CNRS, Astroparticule et Cosmologie, 75013 Paris, France
         \and
             GEPI, Observatoire de Paris, Université PSL, CNRS, 5 Place Jules Janssen, 92190 Meudon, France
         \and
             CNRS, IRAP, 9 Av. colonel Roche, BP 44346, 31028 Toulouse cedex 4, France
             }

   \date{Received ; accepted }
 
  \abstract{
  Massive black hole (BH) mergers will be key targets of future gravitational wave and electromagnetic observational facilities. In order to constrain BH evolution with the information extracted from BH mergers, one must take into account the complex relationship between the population of merging BHs and the global BH population. We analysed the high-resolution cosmological radiation-hydrodynamics simulation \textsc{Obelisk}, run to redshift $z=3.5$, to study the properties of the merging BH population, and its differences with the underlying global BH population in terms of BH and galaxy properties. In post-processing, we calculated dynamical delays between the merger in the simulation at the resolution limit and the actual coalescence well below the resolution scale. We find that merging BHs are hosted in relatively massive galaxies with stellar mass $M_\ast\gtrsim10^9\,M_\odot$. Given that galaxy mass is correlated with other BH and galaxy properties, BH mergers tend to also have a higher total BH mass and higher BH accretion rates than the global population of main BHs. These differences generally disappear if the merger population is compared with a BH population sampled with the same galaxy mass distribution as merger hosts. Galaxy mergers can temporarily boost the BH accretion rate and the host's star formation rate, which can remain active at the BH merger if sub-resolution delays are not taken into account. When dynamical delays are taken into account, the burst has generally faded by the time the BHs merge.
  BH spins are followed self-consistently in the simulation under the effect of accretion and BH mergers. We find that merging BHs have higher spins than the global population, but similar or somewhat lower spins compared to a mass-matched sample. For our sample, mergers tend to decrease the spin of the final BH remnant.
  }
  
   \keywords{ quasars: supermassive black holes -- Galaxy: evolution -- Methods: numerical --
                Gravitational waves
               }

   \maketitle
%

\section{Introduction}

The mergers of black hole (BH) binaries with masses within the $10^4-10^7\,\Msun$ range are key targets for the future space-based Laser Interferometer Space Antenna \citep[LISA,][]{2023LRR....26....2A} and for the proposed missions 
TianQin \citep{2016CQGra..33c5010L} and Taiji \citep{2020IJMPA..3550075R}. 
The gravitational wave (GW) signals from these merging BH binaries are expected to be detected with a signal-to-noise ratio of hundreds up to $z\sim 10$ for mass ratios not too far from unity.
The detection of GWs from these sources is expected to shed light on the astrophysical processes affecting the evolution of BH binaries \citep{2023LRR....26....2A}, perform general relativity tests \citep{2022LRR....25....4A}, and constrain cosmological scenarios \citep{2022arXiv220405434A}.

In order to perform astrophysical inference on how the merging population can inform us about the evolution of BHs and galaxies, it is needed to develop theoretical models to provide guidance. Modelling merging BHs, however, is a complex problem \citep{1980Natur.287..307B} exacerbated by the need to embed their dynamical evolution in a cosmological context \citep{2003ApJ...582..559V}, which requires including additional physics, such as BH formation \citep{2007MNRAS.377.1711S}, their growth, and spin evolution \citep{2005ApJ...620...59S,Berti2008}.

Early work in this area was based on semi-analytic models \citep{2004ApJ...611..623S,2005ApJ...623...23S,Barausse2012}, which have the advantage of rapid computational times that allow for parameter space exploration. Later semi-analytic models have refined various aspects of the implementation of BH dynamics \citep{2019MNRAS.486.4044B,Izquierdo-Villalba2022}. Hydrodynamical cosmological simulations have also been used to study the problem \citep{2016MNRAS.463..870S, 2020MNRAS.491.2301K,Volonteri2020,DeGraf2021,Ni2022}: simulations, in general, are unable to resolve halos as small as in semi-analytical models based on the Extended Press \& Schechter formalism \citep{1993MNRAS.262..627L} and, being computationally expensive, cannot be used to easily investigate variants of fiducial models but have the advantage of treating hydrodynamical processes naturally and of providing spatially resolved information for galaxies. 

Despite the large community effort, the rates and properties of massive BH mergers and their relation to the underlying BH and galaxy populations have large uncertainties in the mass-redshift range of interest for LISA because these galaxies and BHs are poorly probed by observations and because the modelling of BH mergers in a cosmological context is complex. \cite{2023LRR....26....2A} review the current understanding, uncertainties, and theoretical bottlenecks. 

In the following, we study the properties of the BH merger population in the \textsc{Obelisk} simulation \citep{Trebitsch2021} and compare it with the underlying global galaxy and BH populations, with the aim of assessing how and at which level BH evolution can be constrained from BH mergers. \textsc{Obelisk} is a cosmological radiation-hydrodynamical simulation evolving a protocluster down to redshift $\sim 3.5$. This simulation is ideal for our purposes since it has a high resolution ($\sim 35\,\si{\parsec}$) and incorporates detailed models for a wide range of BH physical processes, such as accretion, feedback, spin evolution, and dynamical friction, which are key in order to produce a realistic BH merger population.

In Sect.~\ref{sec:Method}, we summarise the properties of the \textsc{Obelisk} simulation and the identification and selection criteria of galaxies and BH mergers. We also describe our calculation of sub-grid merger delays. We present our results in the subsequent sections -- in Sect.~\ref{sec:cosmic_evol_BHs}, we summarise the growth and evolution of BHs in \textsc{Obelisk}; in Sect.~\ref{subsec:pops_fixed_z} we study the properties of the BH merger population as compared to the full sample of main BHs at a fixed redshift, while in Sect.~\ref{subsec:redshift_evolution} we study the redshift evolution of such properties. Finally, in Sect.~\ref{sec:conclusions}, we conclude and summarise our main results. This paper will be followed by a study of the multi-messenger detectability (GW and electromagnetic) of the \textsc{Obelisk} BH merger sample \citep{Dong-Paez2023b}

\section{Method}
\label{sec:Method}

\subsection{The \textsc{Obelisk} simulation}
\label{subsec:Obelisk}

In this paper, we study the BH population in the \textsc{Obelisk} cosmological simulation \citep{Trebitsch2021}.
\textsc{Obelisk} is a high-resolution ($\Delta x \simeq \SI{35}{\parsec}$), radiation-hydrodynamical simulation, which follows the evolution of 
an overdense region in the \textsc{Horizon-AGN} \citep{Dubois2014c} volume until redshift $z \simeq 3.5$. Below we present a brief summary of the properties of the simulation. For a more detailed description, we refer the reader to \citet{Trebitsch2021}.

The simulation assumes a \textLambda CDM cosmology with the best-fit parameters of the WMAP-7 analysis \citep{Komatsu2011} -- Hubble constant $H_0 = \SI{70.4}{\kilo\meter\per\second\per\mega\parsec}$, dark energy density parameter $\Omega_\Lambda = 0.728$, total matter density parameter $\Omega_\mathrm{m} = 0.272$, baryon density parameter $\Omega_\mathrm{b} = 0.0455$, amplitude of the power spectrum $\sigma_8=0.81$, and spectral index $n_\mathrm{s}=0.967$. The initial conditions for \textsc{Obelisk} consider the particles within a $2.51 \, h^{-1}\,\mathrm{cMpc}$ radius from the centre of the most massive halo in \textsc{Horizon-AGN} at $z=1.97$. The convex hull enclosing these particles in the initial conditions was simulated at high resolution, with a dark matter (DM) mass resolution of $1.2\times10^6\,\Msun$, while the remaining volume of the original $100 \, h^{-1} \, \mathrm{cMpc}$ \textsc{Horizon-AGN} box maintained a lower resolution. 

The \textsc{Obelisk} simulation was run with \textsc{Ramses-RT} \citep{Rosdahl2013,Rosdahl2015}, a radiative transfer hydrodynamical code which builds on the adaptive mesh refinement (AMR) \textsc{Ramses} code \citep{Teyssier2002}. Any cell was refined up to a smallest size of $35\,\si{\parsec}$ if its mass exceeds $8$ times the mass resolution.
The gas is evolved using an unsplit second-order MUSCL-Hancock scheme for the Euler equations, with a minmod limiter to guarantee the total variation diminishing property of the scheme. The simulation assumes an ideal monoatomic gas with adiabatic index $\gamma = 5/3$ and includes gas cooling and heating down to very low temperatures ($50\,\rm K$) with non-equilibrium thermo-chemistry for hydrogen and helium, and contribution to cooling from metals (which ionisation levels are obtained by assuming equilibrium with a standard ultraviolet background) released by supernovae (SNe).
Young stars and BHs are sources of radiation.
 
The stellar population is modelled with $\sim 10^4\,\solarmass$ particles, each representing a population with a \citet{Kroupa2001} initial mass function between $0.1$ and $100 \,\solarmass$. Cells with gas density higher than $5\,\mathrm{H}\,\si{\per\cm\cubed}$ and turbulent Mach number $\mathcal{M}\geq 2$ form stars according to a Schmidt law with a star formation efficiency which depends on $\mathcal{M}$ and the ratio of turbulent to gravitational energy. After $5\,\si{\mega\year}$ from the birth of a star particle, a fraction of $0.2$ the stellar mass explodes as SNe, releasing an energy of $10^{51}\,\si{\erg}$ per explosion to the gas following the numerical implementation from~\cite{kimm&cen14}.

The dust mass fraction of the gas is also followed in the simulation as a passive variable. The model assumes that dust grains of single fixed size and composition are created in SN explosions, that they can accrete from the gas-phase metals, and that they are destroyed by SN explosions and by thermal sputtering.

The BHs in the simulation are seeded with an initial mass of $M_{\bullet,0} = 3\times 10^4 \, \solarmass$ when in a given cell both gas and stars exceed a density threshold of $100 \, \si{H} \, \si{\per\centi\meter\cubed}$ and the gas is Jeans unstable. Additionally, in order to avoid forming multiple BHs in one galaxy, any new BH must not be closer than $50$ comoving kpc to any other already existing BH. The BHs can subsequently grow by gas accretion or BH-BH mergers. Gas accretion is modelled as Bondi-Hoyle-Lyttleton (BHL) accretion,
\begin{equation}
    \dot{M}_\mathrm{BHL} = \frac{4\pi G^2 M_\bullet^2 \bar{\rho}}{\left(\bar{c_s}^2 + \bar{v}_\mathrm{rel}^2\right)^{3/2}},
    \label{eq:BHL_accretion}
\end{equation}
where $G$ is the gravitational constant. The local average gas density, gas sound speed, and the relative velocity of the BH with respect to the gas are denoted by $\bar{\rho}$, $\bar{c_s}$, and $\bar{v}_\mathrm{rel}$. The accretion rate is capped at the Eddington rate
\begin{equation}
    \dot{M}_\mathrm{Edd} = \frac{4\pi GM_\bullet m_{\rm p}}{\varepsilon_{\rm r} \sigma_\mathrm{T} c},
\end{equation}
where $m_{\rm p}$ is the proton mass, $\varepsilon_{\rm r}$ is the radiative efficiency, $\sigma_\mathrm{T}$ is the Thompson cross-section, and $c$ is the speed of light. A fraction $\varepsilon_{\rm r}$ of the mass accreted is radiated, while the remaining $1-\varepsilon_{\rm r}$ is accreted onto the BH, contributing to its mass growth. Defining the Eddington ratio as $f_\mathrm{Edd} = \dot{M}/\dot{M}_\mathrm{Edd}$ (here $\dot{M} = \min{\{\dot{M}_\mathrm{BHL},\dot{M}_\mathrm{Edd}\}}$), for low Eddington ratio sources below $f_\mathrm{Edd,crit} = 0.01$, gas flows are assumed to be radiatively inefficient, in which case the radiative efficiency is further reduced by a factor $f_\mathrm{Edd}/f_\mathrm{Edd,crit}$.

Part of the energy accreted onto a BH is injected into the surrounding gas, following a dual-mode AGN feedback model. At low Eddington ratio $f_\mathrm{Edd}<f_\mathrm{Edd,crit}$, the AGN is in `radio mode', releasing a fraction $\varepsilon_\mathrm{MCAF}$ of the rest-mass accreted energy as kinetic energy in jets aligned with the BH spin angular momentum. Here, $\varepsilon_\mathrm{MCAF}$ is a polynomial fit to the simulations of \citet{McKinney2012} as a function of BH spin, and it assumes Magnetically Chocked Accretion Flows. At higher $f_\mathrm{Edd}\geq f_\mathrm{Edd,crit}$, the AGN is in `quasar mode' and releases isotropically $15\%$ of the accretion luminosity as thermal energy.

Black holes are also allowed to grow by BH-BH mergers. Two BHs are merged in the simulation if their separation becomes smaller than $4\Delta x$ and if they are gravitationally bound in the absence of other bodies. The simulation incorporates subgrid models to account for the unresolved small-scale dynamical friction from both gas and collisionless particles (stars and DM) \citep[following the implementation presented in][]{Dubois2013,Pfister2019}. The inclusion of dynamical friction leads to a natural behaviour of the orbital evolution of BHs in galaxies, and to a natural delay between galaxy mergers and the mergers of the associated BHs. This will be discussed further in Sect.~\ref{subsec:merger_selection}.

The BH spins are self-consistently evolved on the fly via gas accretion and BH-BH mergers \citep[rather than in post-processing,][]{2021MNRAS.501.2531S}. At high accretion rates ($f_\mathrm{Edd}\geq f_\mathrm{Edd,crit}$), the evolution of the spin magnitude $a$ due to gas accretion is calculated according to the following expression \citep{Bardeen1970},
\begin{equation}
    a_{n+1} = \frac{1}{3}\frac{r_\mathrm{isco}^{1/2}}{M_\mathrm{ratio}}\left[4-\left(3\frac{r_\mathrm{isco}}{M_\mathrm{ratio}^2}-2\right)^{1/2}\right],
\end{equation}
where the subscript $n$ denotes the value of the variable at timestep $n$ and $M_\mathrm{ratio}$ is defined as the ratio $M_{\bullet,n+1}/M_{\bullet,n}$. The radius of the innermost stable circular orbit in units of the gravitational radius, $r_\mathrm{isco}$, depends solely on the spin magnitude and the alignment with the disc angular momentum. The direction of the BH spin $\boldsymbol{J}_{{\rm BH},n+1}$ is evolved assuming the direction of the gas angular momentum is conserved from the resolved vicinity of the BH to the unresolved disc $\boldsymbol{J}_{\rm d}$ and is obtained by simply adding $\boldsymbol{J}_{{\rm BH},n+1}=\boldsymbol{J}_{{\rm BH},n}+\boldsymbol{J}_{\rm d}$. The update of the spin magnitude (its co- or counter-rotation with respect to the disc rotation) is decided following the anti-alignment criterion from~\cite{King05}.

At lower accretion rates ($f_\mathrm{Edd}<f_\mathrm{Edd,crit}$), corresponding to radiatively inefficient accretion, rotational energy is assumed to power the radio jets and therefore the magnitude of BH spins decreases as rotational energy is extracted. These BHs are spun down following the polynomial fits in \citet{McKinney2012} with the same procedure for the update of the spin direction as for the $f_\mathrm{Edd}\geq f_\mathrm{Edd,crit}$ case.

Spin also evolves following the coalescence of two BHs. The spin of the remnant BH is modelled using an analytical fit from \citet{Rezzolla2008a},
\begin{equation}
    \boldsymbol{a}_f = \frac{1}{(1+q)^2}\left(\boldsymbol{a}_1 +\boldsymbol{a}_2 q^2 +\boldsymbol{\ell} q \right),
    \label{eq:merger_spin}
\end{equation}
where $\boldsymbol{a}_1$ and $\boldsymbol{a}_2$ are the spins of the primary and secondary respectively, and $q \equiv M_2/M_1$, $M_1$ and $M_2$ being the primary and secondary mass ($M_1\geq M_2$), respectively. We define $\boldsymbol{\ell} \equiv (\boldsymbol{J}-\boldsymbol{J}_\mathrm{gw})/(M_1 M_2)$, where $\boldsymbol{J}$ 
is the orbital angular momentum of the binary when the BHs are widely separated and $\boldsymbol{J}_\mathrm{gw}$ is the angular momentum extracted by gravitational waves. The calculation of $\boldsymbol{\ell}$ is performed as 
\begin{equation}
\begin{split}
\ell = & \frac{s_4}{\left(1+q^2\right)^2}\left(a_1^2 + a_2^2 q^4 + 2 \boldsymbol{a}_1 \cdot \boldsymbol{a}_2 q^2\right)\\ 
& + \frac{s_5\mu + t_0 + 2}{1+q^2}\left(a_1 \cos \phi_1 + a_2 q^2 \cos \phi_2\right) \\
& + 2\sqrt{3} + t_2\mu + t_3\mu^2,
\end{split}
\end{equation}
where $\phi_1$ and $\phi_2$ are the angle of $\boldsymbol{a}_1$ and $\boldsymbol{a}_2$ with $\boldsymbol{\ell}$, and $\mu = q/(1 + q)^2$. The values of the constants are $s_4 = -0.129$, $s_5 = -0.384$, $t_0 = -2.686$, $t_2 = -3.454$, and $t_3 = 2.353$. The direction of $\boldsymbol{\ell}$ is assumed to be aligned with the orbital angular momentum of the BH binary. The value of the spin is used to determine the efficiency of the energy injection into jets in the radio mode, and the BH radiative efficiency, as
\begin{equation}
    \varepsilon_{\rm r} = 1-\sqrt{1-\frac{2}{3r_\mathrm{isco}}}.
\end{equation}

\subsection{Galaxy catalogues and BH-galaxy matching}

We identified galaxies and their DM haloes together using a version of the \textsc{AdaptaHOP} algorithm \citep{Aubert2004, Tweed2009} designed to work on both stars and DM particles. Substructures were identified using the most massive sub-maximum method \citep[see details in][]{Trebitsch2021}. In the initial selection, structures were only selected if the total number of particles (stars + DM) exceeds 100.
As \textsc{Obelisk} is a resimulation of \textsc{Horizon-AGN}, some structures can in principle be polluted by low-resolution (massive) DM particles. To avoid uncontrolled dynamical effects from poorly resolved structures, we excluded haloes that contain any low-resolution particles, effectively restricting ourselves to perfectly un-contaminated haloes. 

\textsc{Obelisk} is a high-resolution simulation designed to study an overdense region at high redshift, leading to high merger rates and disturbed morphologies. As such, the distinction between a galaxy and a star-forming clump can be difficult. In this work, we only considered as galaxies those structures that have at least 100 star particles and 100 DM particles. While this removes the spurious, DM-free stellar clumps, our galaxy catalogue might still contain structures that are DM-deficient. These can originate for instance from mergers.
Similarly, the definition of the galaxy centre can be affected by the presence of these clumps, which is not just a numerical artefact: the meaning of the centre for a clumpy, high-redshift galaxy is not trivial to define. As such, our actual catalogue includes multiple definitions for centres. On top of the centres of mass (for stars, DM, and both), we followed the approach used in the \textsc{New-Horizon} simulation \citep{Dubois2021} and chose as our fiducial centre the position of the density peak (for stars, DM, and both) determined recursively using a shrinking sphere approach.

We define the main BH of a given galaxy to be the most massive BH located within the half-mass radius $r_{50}$, or within $4\Delta x$ if $r_{50}<4\Delta x$. The BHs that are not assigned to any galaxy as main BHs can be assigned as satellite BHs to the highest stellar mass galaxy within whose $2r_{50}$ the BH is located, or within $4\Delta x$ if $2r_{50}<4\Delta x$.

Galaxy properties were obtained from the simulation snapshots. The BH properties are stored every coarse timestep, which occurs on a much shorter time scale than simulation snapshots. For comparison, there are a total of $205$ snapshots, for $10076$ coarse timesteps.
Galaxy star formation rates were averaged on a time window of $\SI{10}{\mega\year}$, while all other quantities were taken to be their instantaneous value.

\subsection{Selection of BH mergers}
\label{subsec:merger_selection}

All BHs are assigned a unique ID number at seeding. Black holes are only removed from the simulation following a BH-BH merger, when two BHs approach each other at a distance smaller than $4 \Delta x$ and are gravitationally bound in vacuum. The least massive BH or secondary disappears from the simulation, while the most massive BH or primary preserves its ID number, gains the mass of the secondary, and updates its spin following the coalescence. We identified these numerical mergers by searching for BHs that disappear from the simulation. These missing BHs can be identified as numerical merger secondaries. The corresponding primaries can be found by searching for more massive BHs at a distance smaller than $4 \Delta x$ in the coarse timestep prior to the merger. 

A BH merger was assigned to a galaxy if it is found within $r_{50}$ of the galaxy in the snapshot immediately after the merger. In the following, we restrict our analysis to BH mergers occurring within $r_{50}$ of a galaxy, eliminating possible spurious cases. Mergers occurring in galaxies represent $878$ out of the total of $1543$ numerical mergers.

While these numerical mergers occur when the separation of the two BHs is about $140\,\si{\parsec}$, the physical merger will take place when the separation is of the order of their gravitational radius, $10^{-6}  (M_{\rm \bullet}/10^7 \,  \Msun)\, {\rm pc}$. The physical merger is delayed compared to the numerical merger by dynamical processes. We define `delayed mergers' as the outcome of adding delays calculated in post-processing. We followed the methodology described in \cite{Volonteri2020}, which we briefly summarise here. 

First, we added a dynamical friction phase from the position where the BHs are merged at the numerical merger down to the centre of the host galaxy, at which point we assume that a binary forms. This dynamical friction delay was computed as the dynamical friction timescale for a massive object in an isothermal sphere, considering only the stellar component of the galaxy and including a factor $0.3$ to account for typical orbits being non-circular:
\begin{equation}
t_{\rm df}=0.67 \left(\frac{d}{4\, {\rm kpc}}\right)^2\left(\frac{\sigma_\ast}{100 \, {\rm km\, s^{-1}}}\right)\left(\frac{M_\bullet}{10^8 \, M_{\odot}}\right)^{-1}\Lambda^{-1}\, {\rm Gyr}\, ,
\label{eq:tdf}
\end{equation}
where $M_\bullet$ is the black hole mass, $\sigma_\ast$ is the central stellar velocity dispersion approximated as $(0.25GM_\ast/r_{50})^{1/2}$, $\Lambda=\ln(1+M_\ast/M_\bullet)$, with $M_\ast$ being the total stellar mass of the galaxy hosting the BH at the output before the numerical merger, and $d$ its distance from the centre. 
That the stellar component of the galaxy is the main driver for dynamical friction has been confirmed by \cite{2020ApJ...896..113L}, \cite{2020ApJ...905..123L}, and \cite{2022MNRAS.510..531C}.  We calculated the sinking time of both BHs in a pair and take the longer of the two, which is normally associated with $M_2$.

We also accounted for the shrinking of the binary until coalescence via stellar hardening (assuming an inner power-law density profile with slope $-2$), viscous torques in a circumbinary disc, and emission of gravitational waves, following \cite{2015MNRAS.454L..66S} and \cite{2015MNRAS.448.3603D}. The binary evolution timescale to coalescence was taken to be the minimum between the two following equations:
\begin{equation}
t_{\rm bin,h}=1.5 \times 10^{-3} \left(\frac{\sigma_{\rm inf}}{\rm 100 \,km\,s^{-1}}\right) \left(\frac{\rho_{\rm inf}}{10^6 \, \Msun \,{\rm pc}^{-3}}\right)^{-1}\left(\frac{a_{\rm gw}}{10^{-3}\,\rm{pc}}\right)^{-1} \, {\rm Gyr}\, , 
\label{eq:tbinh}
\end{equation}
and
\begin{equation}
t_{\rm bin,d}=1.5\times10^{-2} \,\varepsilon_{0.1} \,f_{\rm Edd}^{-1} \frac{q}{(1+q)^2}\ln\left(\frac{a_{\rm i}}{a_{\rm c}}\right)  \, {\rm Gyr}\, . 
\label{eq:tbind}
\end{equation}
In Eq.~\ref{eq:tbinh}, $\sigma_{\rm inf}$ and $\rho_{\rm inf}$ are the velocity dispersion and stellar density at the sphere of influence, defined as the sphere containing twice the binary mass in stars, and $a_{\rm gw}$ is the separation at which the binary spends most of the time \citep[see][]{2015MNRAS.454L..66S}. In Eq.~\ref{eq:tbind} $\varepsilon_{0.1}$ is the radiative efficiency normalised to 0.1. We followed \cite{2015MNRAS.448.3603D} in selecting $a_{\rm i}=G M_{12}/2\sigma_\ast^2$ and $a_{\rm c}=1.9\times10^{-3} (M_{12}/10^8 \, M_{\odot})^{3/4}\, {\rm pc}$, where $M_{12}$ is the total mass of the binary. 

The longest delay is generally caused by small-scale dynamical friction, which has a median of $t_\mathrm{df}$ is $1.3\,\si{\giga\year}$ with an interquartile scatter of $1.6\,\si{\giga\year}$. If we restrict the analysis to MBHs whose dynamical friction time elapses before $z=3.5$ (the end of the simulation) and which therefore form a binary during the simulated time-frame, the value decreases to $0.3\,\si{\giga\year}$ with the same scatter. For these MBHs we can further calculate binary evolution timescales due to either hardening+GWs or migration+GWs, whichever is shorter. The timescale $t_{\rm bin,h}$ has a median value of $0.13\,\si{\mega\year}$, and $t_{\rm  bin,d}$ of $0.38\,\si{\giga\year}$ (both with large scatter, about twice the median value). The hardening timescales are very short as a consequence of high redshift galaxies being compact \citep{2017JPhCS.840a2025M}. The delay timescales, under our assumptions, are therefore dominated by small-scale dynamical friction. We have to restrict our analysis of $t_{\rm bin,h}$ and $t_{\rm bin,d}$ to the sub-sample of mergers whose dynamical friction time elapses before the end of the simulation, as no further information on the evolution of the mergers' host galaxy is available after the simulation has ended. 
A significant fraction of delayed mergers is predicted to occur at $z<3.5$, after the final redshift to which \textsc{Obelisk} has been run, and therefore cannot be modelled in this way.
Since a large fraction of delayed mergers results from numerical mergers of BHs close to their seed mass, our estimated delay times are dependent on our choice of the seed mass.

While this post-processing procedure allows one to calculate when `delayed mergers' happen, the evolution of the binary mass ratio is poorly defined and a variety of outcomes is possible \citep{Siwek2020}. A further complication is that in the most massive galaxies of \textsc{Obelisk} multiple BH mergers occur between numerical and delayed mergers, further complicating any approach on how to post-process the evolution of the mass-ratio of a binary. Similarly, the merger-induced spin change is only introduced at the numerical merger, while the spin of delayed mergers is further modulated by the intervening accretion and mergers.  When a quantity cannot be reasonably predicted in post-processing we refrain from using delayed mergers and consider only numerical mergers. Otherwise, we use delayed mergers under the assumption that the total BH mass at the time of the delayed merger corresponds to the BH mass of the merger remnant in \textsc{Obelisk} at that time. 
We considered both numerical mergers (in which the merger takes place at unrealistically large distances but all parameters are consistently evolved) and delayed mergers (where the evolution is post-processed with simplified equations) to bracket our uncertainty.

We measured the primary and secondary BH properties at the coarse timestep immediately before the numerical merger. The properties of their host galaxy were measured at the simulation snapshot prior to the merger. Similarly, the properties of the numerical remnant BH were measured at the coarse timestep immediately after the merger, while its host galaxy properties were measured at the simulation snapshot after the merger. For the delayed merger remnants, BH and galaxy properties were measured at the coarse timestep and snapshot after the delayed merger, respectively.

To trace the galaxy merger that gave rise to a BH merger we followed the BHs backwards in time until we found that the two BHs are assigned to different galaxies, and we considered this the time of the galaxy merger. The consequences of a merger on gas and stellar properties, for example, a star formation or accretion burst, occur after a few dynamical times \citep[after a close pericentric passage][]{2015MNRAS.447.2123C}, therefore merger-induced star formation rate (SFR) or $f_\mathrm{Edd}$ boosts will occur after the galaxy merger. 

\section{The cosmic evolution of BHs in \textsc{Obelisk}}
\label{sec:cosmic_evol_BHs}

\cite{Trebitsch2021} presents the general trends of the BH population in \textsc{Obelisk} as well as a comparison with observations. The largest BH masses found at the end of the simulation, $z=3.5$, are $\sim 10^8 \, \Msun$. \textsc{Obelisk} is the simulation of a protocluster, that is to say a biased region, but not an extreme overdensity. As a consequence, we do not find any BH with mass $M_\bullet\gtrsim10^9\,\solarmass$, which are typical of bright quasars found in observations at $z\sim 6-7$ \citep[e.g.][]{Banados2018,Wang2021,Yang2021}. As mentioned in \citet{Trebitsch2021}, these quasars are rare, with number densities of $\sim10^{-9}\, \mathrm{cMpc}^{-3}$ \citep{Wang2019} and are plausibly hosted in halos with mass $10^{12}-10^{13} \,\Msun$ at $z\sim 6$ \citep{2001AJ....122.2833F}. These BHs would only be expected to occur for simulation volumes much larger than \textsc{Horizon-AGN}, the parent simulation of \textsc{Obelisk}. The BHs in \textsc{Obelisk} are not rare and possibly exceptional BHs but are representative of the general population. One should keep in mind, however, that in a protocluster environment galaxy evolution is accelerated, with respect to galaxies in the field. 

In the following, we discuss the evolution of the BHs in the simulation. The evolution of BHs is driven by the complex interplay between galactic and BH processes. We highlight some common trends that describe the main characteristics of BH evolution. 

\begin{figure}
	\includegraphics[width=\columnwidth]{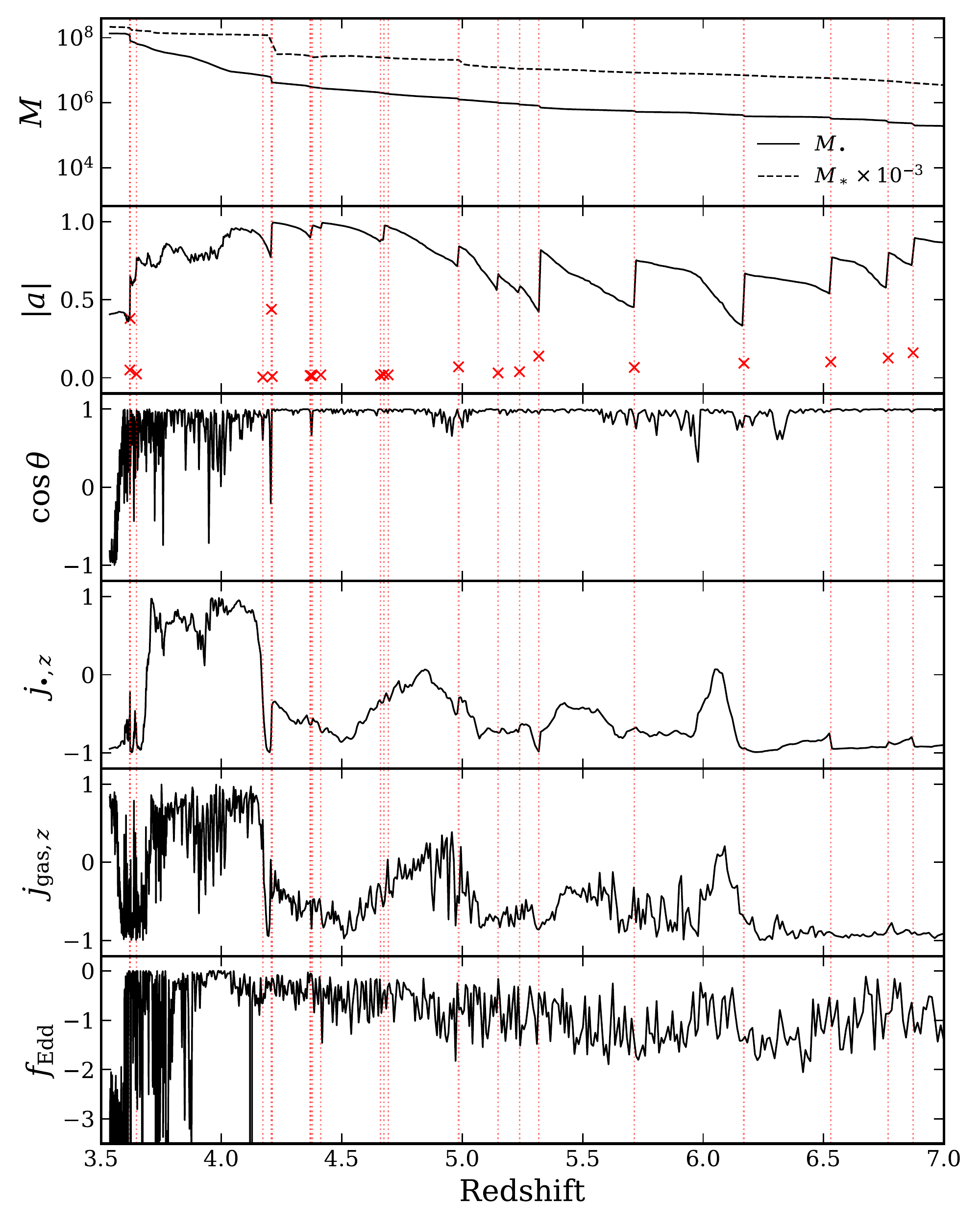}
    \caption{Cosmic evolution of BH 2, the second most massive BH at redshift $\sim 3.5$, which reaches a mass $M_\bullet \sim 10^8\,\solarmass$ at the end of the simulation. The BH mass $M_\bullet$ (top panel, solid line), galaxy mass $M_\ast$ (dashed line), the spin magnitude $|a|$, the cosine of the angle $\theta$ between the spin direction $\boldsymbol{j}_\bullet$ and the gas angular momentum $\boldsymbol{j}_\mathrm{gas}$, the $z$-components of $\boldsymbol{j}_\bullet$ and $\boldsymbol{j}_\mathrm{gas}$, and the Eddington ratio $f_\mathrm{Edd}$, are plotted against the cosmological redshift. Vertical red dotted lines indicate the time of a numerical BH merger, with the vertical position of the crosses showing its mass ratio $0<q\leq1$.}
    \label{fig:BH_history_massive}
\end{figure}
\begin{figure}
	\includegraphics[width=\columnwidth]{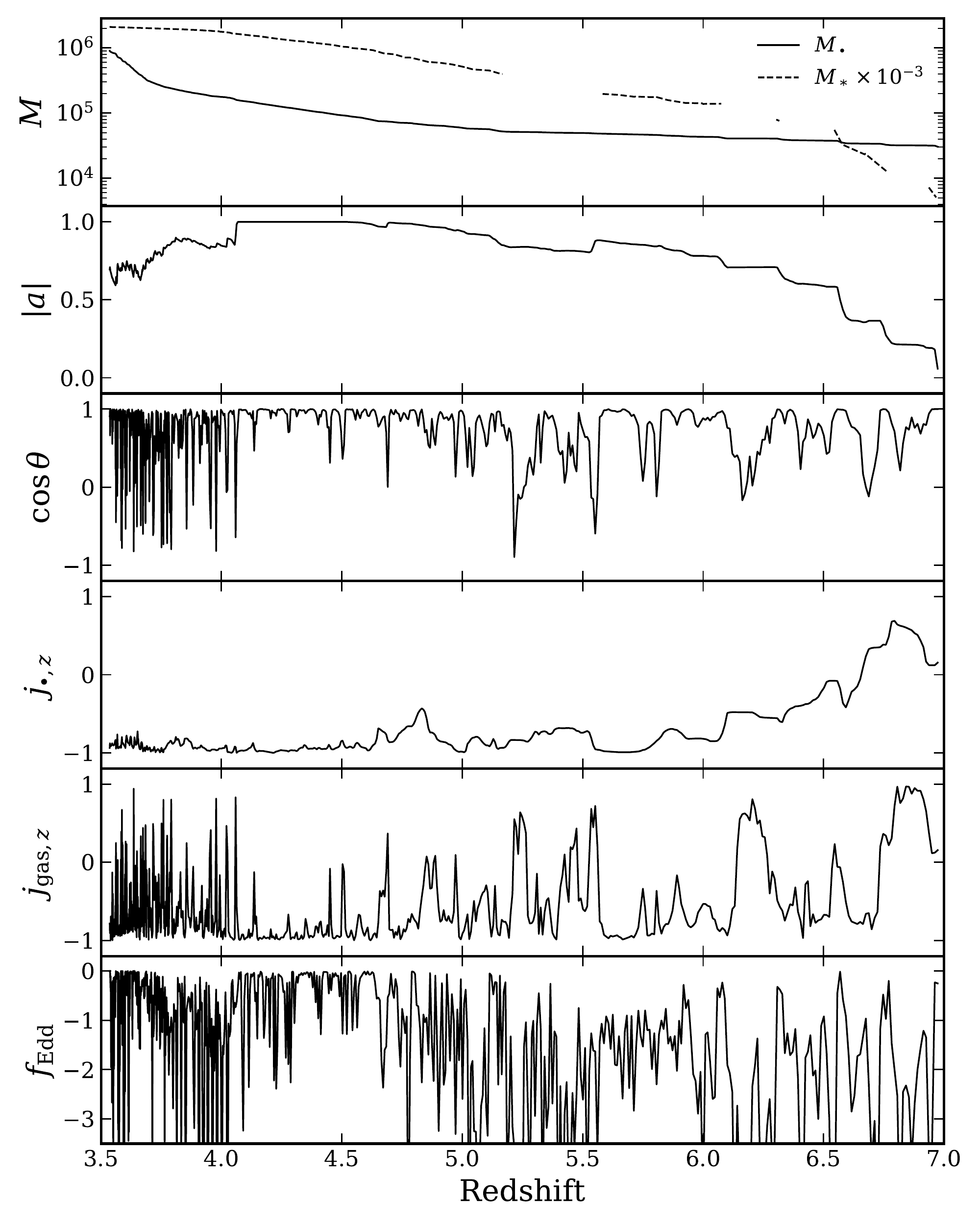}
    \caption{Similar to Fig.~\ref{fig:BH_history_massive}, for BH 2108, which reaches $M_\bullet \sim 10^6\,\solarmass$ at the end of the simulation ($z\sim3.5$). At the times where $M_\ast$ is absent from the figure, the BH is not assigned to any galaxy.
    }
    \label{fig:BH_history_intermediate}
\end{figure}
The evolution of a set of properties for two BHs with final masses $M_\bullet\sim10^8\,\solarmass$ and $M_\bullet\sim10^6\,\solarmass$ at the end of the simulation $z\sim 3.5$ is shown in Figs.~\ref{fig:BH_history_massive} and~\ref{fig:BH_history_intermediate}. We show the BH and galaxy masses, the magnitude of the BH spin, the alignment between the BH and gas angular momentum, the $z$-component (in the frame of the cartesian axis of the simulation box) of the BH and gas angular momentum, and the accretion rate in Eddington units. We denote BH merger events by red dashed vertical lines, and indicate the value of the mass ratio $q$ with crosses.

Black holes grow through extended periods of gas accretion and BH merger events. Black hole mergers tend to be clustered in time, as can be seen in Fig.~\ref{fig:BH_history_massive}. A galaxy merger can bring in multiple BHs from the satellite galaxy, some of which are susceptible to merging with the main BH in the larger galaxy on similar timescales\footnote{The corresponding delayed mergers will however not necessarily occur at a concurrent time.}, modulated by dynamical friction. During the periods of gas accretion, the transfer of angular momentum from the gas to the BH aligns the spin with the gas angular momentum. Similarly, BH merger events modify the spin according to the binary angular momentum and the individual binary BH spins.

Gas accretion onto a BH is regulated by the properties of the host galaxies. Low-mass galaxies lack a well-defined disc, and their shallow gravitational potential makes their gas susceptible to being strongly perturbed or ejected by SN feedback~\citep{Dubois2015,habouzitetal17,Lapiner2021}. The BHs will often accrete very inefficiently from gas with erratic angular momentum, leading to small spin values in the early phase of their evolution~\citep{Dubois2014b}.

It is only for $M_\ast \gtrsim 10^9\,\solarmass$ that galaxies stabilise dynamically and form a disc or proto-disc, allowing BHs to accrete coherently at high rates. This explains why the BH in Fig.~\ref{fig:BH_history_massive} grows efficiently already at $z = 7$, while the less massive BH in Fig.~\ref{fig:BH_history_intermediate} only starts growing efficiently at $z\sim 6$. The BHs in this regime grow rapidly and tend to attain almost maximal spin values. Even in the case of a sharp merger-induced spin decrease, the BH is able to recover swiftly a high spin value via strongly coherent gas accretion. We also note that more massive BHs will accrete faster given our BHL accretion model in which at fixed nuclear gas properties $f_\mathrm{Edd} \propto M_\bullet$ (see Eq.~\ref{eq:BHL_accretion}).

For larger $M_\ast \gtrsim 10^{11}\,\solarmass$, corresponding to $M_\bullet\gtrsim 10^8\,\solarmass$, the availability of gas declines sharply, and the remaining gas becomes hot and turbulent. This decreases the BH accretion rate and the coherence of the accreted gas angular momentum. In this high-mass regime, BH mergers tend to be numerous and dominate the BH mass growth and its spin evolution. These considerations explain the decline in spin and $f_\mathrm{Edd}$ at $z\lesssim 4$ for the BH in Fig.~\ref{fig:BH_history_massive}.

In summary, the BH accretion rate generally increases with the host galaxy mass, with cutoffs at the very low-mass and high-mass ends. These results are in agreement with previous findings in terms of the modulation of the BH accretion rate \citep{Dubois2015,habouzitetal17,2017MNRAS.465...32B,2017MNRAS.472L.109A,Trebitsch2021,Lapiner2021,Beckmann2022,byrneetal23} and of the processes driving spin evolution \citep{Dubois2014a,Dubois2014b,2019MNRAS.490.4133B}.

\section{The population of merging BHs compared to the global BH population in \textsc{Obelisk}}
\label{sec:merger_population}

In this section, we study the general properties of the population of merging BHs and their host galaxies and compare them to the global population of BHs.  It is important to assess at what level the merging population is biased if BH mergers are used to make inferences about BH evolution.

Almost all the BH mergers in our sample are mergers between main BHs. These BH mergers tend to follow the merger of two galaxies. After a galaxy merger, the BHs from each galaxy will tend to sink to the centre of the new potential, and may eventually coalesce if the orbital decay is effective at all scales.
We recall that two BHs are merged numerically in the simulation if the separation becomes shorter than $4\Delta x$ and the BH-BH relative velocity is smaller than the binary escape velocity, as described in Sect.~\ref{subsec:Obelisk}.  We refer to this set of systems as `numerical mergers'. 

Moreover, if the post-processed sub-grid delays are small enough, the system will coalesce before the end of the simulation. We denote this set of systems as `delayed mergers' (see Sect.~\ref{subsec:merger_selection} for definitions and details).
Not all numerical mergers lead to a delayed merger. When delays are included only $\sim 1/6$ of numerical mergers are able to coalesce before the end of the simulation at $z=3.49$, corresponding to an age of the Universe of $1.85\,\si{\giga\year}$. The post-processed small-scale dynamical friction calculated at the numerical merger is shorter than the age of the Universe today for a fraction $\sim 7/8$ of numerical mergers.

As mentioned, the population of merging BHs and associated galaxies is not necessarily representative of the global population of BHs and galaxies. Quite the contrary, massive galaxies residing in dense environments are expected to experience a larger number of interactions and mergers. Any such difference between merger hosts and the global galaxy population will also result in differences between the properties of merging BHs and the global BH population since BH and host properties are correlated. Galaxy mergers themselves might also perturb the properties of galaxies and BHs, for instance producing a star-formation burst or enhancing temporarily the BH accretion rate. 

Numerical mergers do not necessarily happen concurrently with galaxy mergers, as the BHs' orbits have to decay under the effect of dynamical friction, and we recall that to avoid spurious BH mergers we restrict our analysis to numerical mergers located within $r_{50}$ of a galaxy. Therefore the BHs have to reach the centre of the merger remnants. In \textsc{Obelisk} the median time elapsed between the galaxy merger and the BH numerical merger is $110\,\si{\mega\year}$, and $70\,\si{\giga\year}$ if we restrict the analysis to $3.5<z<4$. 
As a reference, \cite{2016A&A...592A..62G} find that both SFRs and BH accretion rates are increased after the first close encounter ($<10$~kpc) and remain elevated for a few hundred of Myr. Numerical mergers may therefore be biased towards galaxy-merger-induced features, such as an increase in SFR or BH accretion rate \citep[see also][]{Volonteri2020,DeGraf2021}

The time delays between numerical mergers and their corresponding potential delayed mergers could in principle re-normalise such features, unless the galaxy merger rate is high, which is expected in the high-redshift Universe \citep[see the discussion in][]{Volonteri2020}. The delay timescales can be very long (for example the median $t_\mathrm{df}$ is $1.3\,\si{\giga\year}$ with an interquartile scatter of $1.6\,\si{\giga\year}$), and they depend on the properties of the BHs and their environments. Consequently, the population of delayed mergers -- which are able to evolve rapidly enough so as to coalesce before the end of the simulation -- will present differences with respect to that of numerical mergers, but also with respect to the general population.

\subsection{BH mergers at $3.5<z<4$}
\label{subsec:pops_fixed_z}

In this section, we analyse the differences in galaxy mass, BH mass, BH accretion rate, galaxy star formation rate, and BH spin between BH merger and the global main BH population. In order to control any possible degeneracies of BH and host properties with the cosmic evolution of BHs and galaxies, we here restrict our analysis to the redshift range $z=3.5-4$.

\subsubsection{Galaxy and BH mass, and BH accretion rate}
\label{subsubsec:Mstar_MBH_and_fEdd}
\begin{figure*}
    \includegraphics[width=\textwidth]{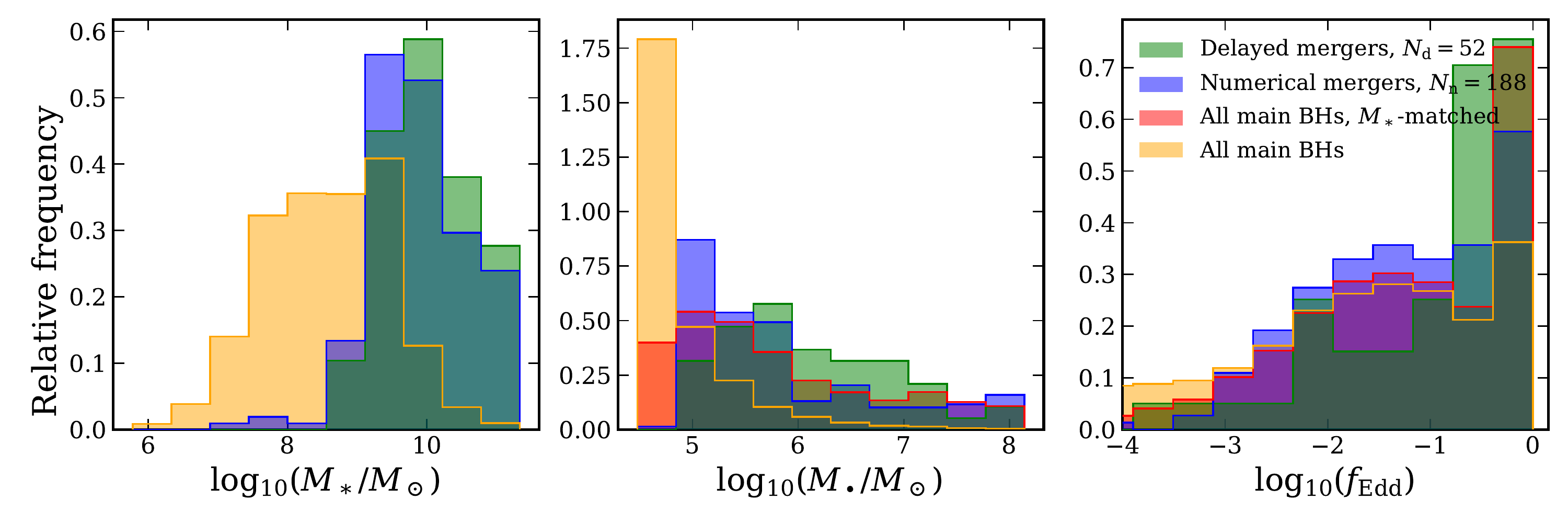}
    \caption{Distribution of host galaxy stellar mass (left panel), BH mass (middle panel), and Eddington ratio (right panel), for delayed merger remnants, in green, numerical merger remnants, in blue, and for the global main 
    BH population, in orange, at redshift $3.5<z<4$. For $M_\bullet$ and $f_\mathrm{Edd}$, we also show in red the global main BH distributions weighted by a factor $\mathrm{pdf}_\mathrm{NM}(\log_{10}M_\ast)/\mathrm{pdf}_\mathrm{all}(\log_{10}M_\ast)$, where $\mathrm{pdf}_\mathrm{NM}(\log_{10}M_\ast)$ and $\mathrm{pdf}_\mathrm{all}(\log_{10}M_\ast)$ are respectively the distribution of $\log_{10}M_\ast$ for numerical merger remnants and all galaxies, as shown in the left panel. In this way, we build a sample of the global main BHs population that matches the $M_\ast$ distribution of merging BHs. Merging BHs have similar properties when compared to a sample matched in $M_\ast$, but have higher masses and Eddington ratios when compared to the full population of BHs.}
    \label{fig:merging_BH_distr}
\end{figure*}
In Fig.~\ref{fig:merging_BH_distr}, we show the distribution of galaxy stellar mass, BH mass, and Eddington ratio for the population of delayed and numerical merger remnants and for the underlying global population of main BHs in galaxies in the redshift range $z=3.5-4$. We find that BH mergers tend to occur in galaxies with higher stellar mass in comparison to the global population of BH hosts -- BH mergers are generally hosted by galaxies with $M_\ast > 10^9\, M_\odot$. More massive galaxies tend to experience a larger number of galaxy mergers, each potentially sourcing BH mergers. More massive galaxies can also bring the BHs to the galactic centre more effectively than dwarf galaxies, which tend to have longer dynamical times due to the presence of DM cores, low densities, and more irregular density profiles \citep{Tremmel2015,Tamfal2018,Bellovary2019,Pfister2019}. Finally, many low-mass galaxies simply don't host BHs, therefore the merger of two galaxies in which only one, or none, hosts a BH does not lead to a BH merger. In \textsc{Obelisk} the fraction of galaxies hosting a BH as a function of galaxy mass, generally referred to as the occupation fraction, reaches 50\% at $M_\ast\sim 10^9\,\Msun$, and unity at $M_\ast \lesssim 10^{10}\,\Msun$, for $3.5<z<7$. This reduces the fraction of galaxy mergers that result in a BH merger for $M_\ast<10^9\,\Msun$.

Since galaxy mass is correlated with other BH and galaxy properties, BH mergers and BH merger hosts will present differences in other properties compared to the global population simply because BH mergers tend to reside in more massive galaxies. If instead we want to analyse the differences between BH mergers and the global main BH population at fixed $M_\ast$ distribution, it is often useful to compare the BH merger population instead with a galaxy-mass-matched sample, in which the global population of BHs is sampled with the same $M_\ast$ distribution as BH mergers hosts \citep[as done in e.g.][]{DeGraf2021}. This can be done by weighting the global main BH distribution by a factor $\mathrm{pdf}_\mathrm{M}(\log_{10}M_\ast)/\mathrm{pdf}_\mathrm{all}(\log_{10}M_\ast)$. Here, $\mathrm{pdf}_\mathrm{M}(\log_{10}M_\ast)$ and $\mathrm{pdf}_\mathrm{all}(\log_{10}M_\ast)$ are the probability density functions of $\log_{10}M_\ast$ for merger remnants and all galaxies respectively. The distributions are obtained by interpolating the histograms shown in the left panel of Fig.~\ref{fig:merging_BH_distr}. These global main BH $M_\ast$-matched distributions are shown with red histograms in Fig.~\ref{fig:merging_BH_distr}. Although in this section we show the global mass-matched population using the $M_\ast$ distribution of numerical mergers, we note that the $M_\ast$ distributions of numerical and delayed mergers are qualitatively similar and produce similar mass-matched populations.

\begin{figure}
	\includegraphics[width=\columnwidth]{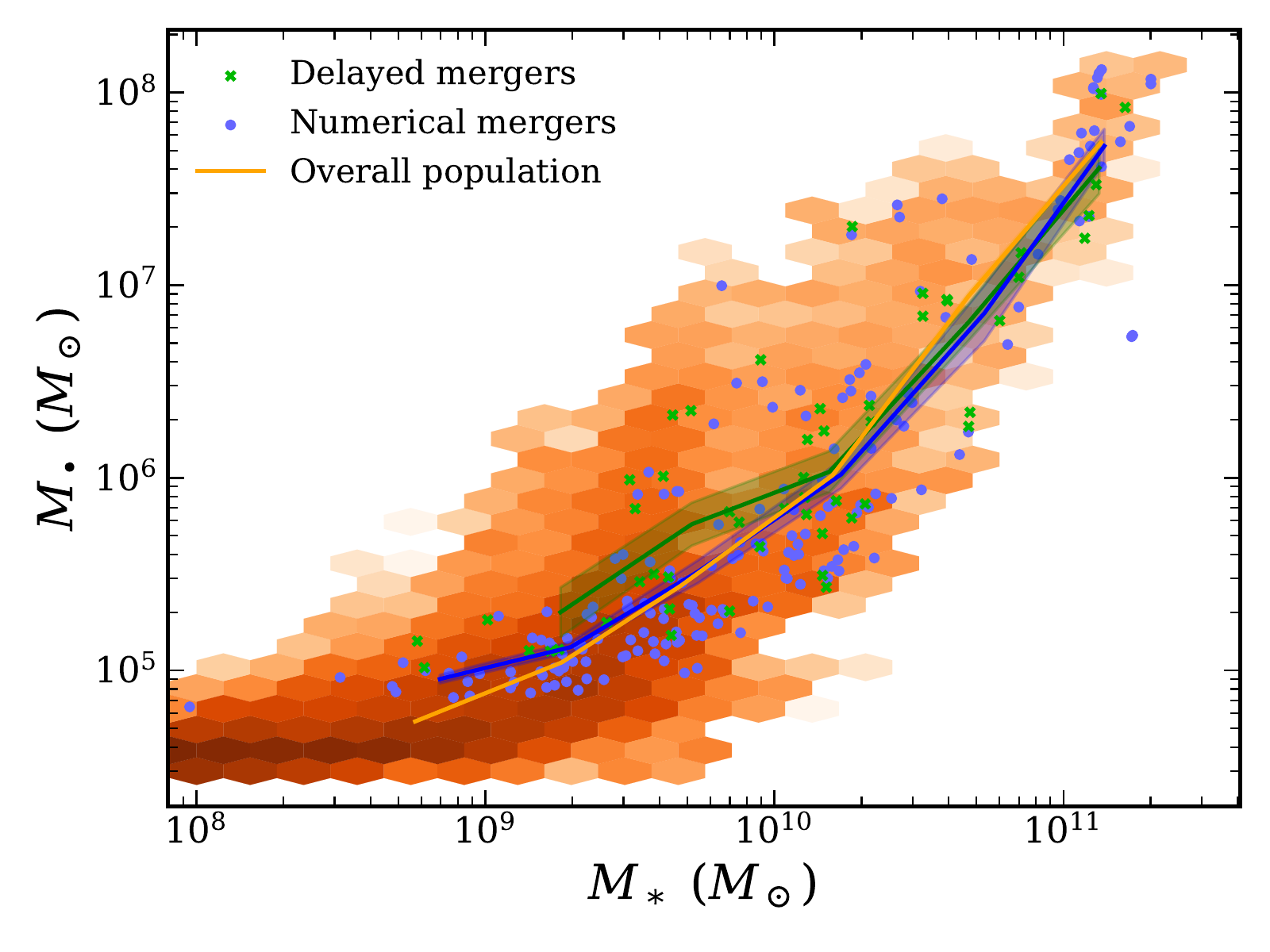}
    \caption{Black hole mass against the host galaxy stellar mass $M_\ast$ for delayed merger remnants (green crosses), numerical merger remnants (blue points), and the global population of main BHs (orange hexagons), at redshift $3.5<z<4$. For the hexagons, darker colours correspond to a higher density of points, on a logarithmic scale. 
    The geometric mean of $M_\bullet$ in several bins of $M_\ast$ is shown by the solid lines for the three samples of BHs, with shaded regions corresponding to the $1\sigma$ error in the mean. The relation for numerical mergers is very similar to that of the full BH population, while the relation for delayed mergers is slightly shifted to higher BH masses in galaxies with mass $M_\ast=10^9-10^{10} \, M_\odot$.
    }
    \label{fig:MBH_vs_Mgal_delay}
\end{figure}

The masses of BH merger remnants are also larger than for the global BH population,
since more massive BHs tend to reside in more massive galaxies. The correlation between galaxy and BH mass is shown in Fig.~\ref{fig:MBH_vs_Mgal_delay}. The relation for numerical mergers follows closely that of the global population. This is consistent with previous findings \citep{Volonteri2020,DeGraf2021,Ni2022}. The flattening of the $M_\ast-M_\bullet$ relation at the low-mass end is a consequence of \textsc{Obelisk}'s SN feedback model, which greatly reduces the efficiency of BH growth at low masses. The asymptotic value of the relation is set by our seeding prescription, which in our case results in seed BHs with a monochromatic initial mass function with $M_{\bullet,0} = \SI{3e4}{\solarmass}$ (see Sect.~\ref{subsec:Obelisk} for more details). For numerical mergers, the higher $M_\bullet$ distributions are consistent with the global $M_\ast$-matched distribution

Similarly, we find that $f_\mathrm{Edd}$ tends to be higher for merger remnants compared to the global main BH population. This difference is consistent with numerical mergers having a higher $M_\ast$ distribution since $f_\mathrm{Edd}$ is correlated with $M_\ast$. As discussed in Sect.~\ref{sec:cosmic_evol_BHs}, $f_\mathrm{Edd}$ generally increases with $M_\bullet$ and $M_\ast$, with cutoffs at very low and high masses. 
The $f_\mathrm{Edd}$ distribution of numerical merger remnants is comparable to that of a galaxy-mass-matched global BH distribution, therefore a merger-induced increase in the BH accretion rate is not very significant in the sample at this redshift (but see Sect.~\ref{subsec:redshift_evolution} below for higher redshift). This is in general agreement with other studies \citep{2018MNRAS.476.2801M,mcalpine20,Smethurst22}, which show that although galaxy mergers induce some BH growth, such growth is not dominant globally.

The distributions of delayed mergers also show differences with respect to the numerical mergers taking place coevally. The population of delayed mergers and numerical mergers occurring at a similar redshift do not represent the same events, since delayed mergers correspond to binaries from earlier numerical mergers that have evolved and coalesced. For our set of numerical mergers, delay timescales are generally long ($\sim 0.1-10 \ \si{\giga\year}$), and so the growth from the numerical merger to the eventual delayed merger can be very significant. In fact, most binaries start with a mass close to the seed mass and gain most of their mass during the post-processed delay time.

In Fig.~\ref{fig:merging_BH_distr} we observe that $M_\ast$ and $M_\bullet$ are both higher on average for delayed mergers compared to numerical mergers due to two reasons. Firstly, any delayed merger occurring at a given time generally corresponds to a galaxy merger occurring at an earlier time, with a significant delay between the two. Early galaxy mergers will leave more time for the dynamical evolution of the BHs and are more likely to result in a delayed merger. Secondly, the merger can directly boost BH growth. The BHs receive a mass boost around the time of the numerical merger since (i) the gas brought in by the galaxy merger can become available for accretion (ii) the two BHs are converted into a single BH in the simulation, artificially boosting the mass and BHL accretion rate of the BHs. We note that this second effect is purely numerical, and thus must be taken with caution.

Since more massive BHs accrete faster in the BHL model, this merger-driven mass gain boosts the subsequent growth in a cumulative fashion \citep[see Appendix A in][]{Volonteri2020}. As seen in Fig.~\ref{fig:MBH_vs_Mgal_delay}, delayed merger remnants tend to be more massive with respect to their host galaxy compared to numerical mergers, especially for mergers with short delay times. However, the average is still well within the scatter of the global relation. 

Additionally, note that only $\sim 1/6$ of numerical mergers become delayed mergers before the end of the simulation. 
The dynamical friction timescale for the formation of a binary is generally dominated by the mass of the secondary BH (see Eq.~\ref{eq:tdf}), thus delayed mergers favour numerical mergers with $q$ initially close to unity. 
Also, numerical mergers occurring close to the end of the simulation require very short delay times in order to be considered delayed mergers, and therefore few of the late numerical mergers are counted as delayed mergers. These two effects suppress the merger of BHs with initially very high masses since the most massive BHs naturally appear only at late times and tend to have been involved in low-$q$ mergers \citep[e.g.][]{Barausse2012}. An illustration of this is that there are no delayed mergers in our sample with an initial (pre-delay) mass larger than $2\times 10^7\solarmass$. 

In summary, as observed in Fig.~\ref{fig:merging_BH_distr}, the delayed BH merger remnant population is more massive on average compared to the coeval numerical remnant population, while having a smaller population of very high mass BHs. The distribution of $M_\ast$ is also consequently characterised by somewhat larger masses.
These results are in qualitative agreement with previous findings \citep{Volonteri2020,DeGraf2021,Ni2022,Izquierdo-Villalba2022}.
The $f_\mathrm{Edd}$ distributions of delayed and numerical mergers are generally similar.

An important caveat to note is that the simulation is stopped at $z \sim 3.5$ which corresponds to a time of $\sim \SI{1.8}{\giga\year}$, while the bulk of the numerical mergers occurs at $z \lesssim 7$, 
which corresponds to a time $\gtrsim \SI{0.8}{\giga\year}$. This limits the maximum delay time to $\sim \SI{1}{\giga\year}$ and biases the delayed merger distribution towards those originating at early times, which naturally involve BHs with a mass close to the seed mass. Given that $\sim 60\%$ of mergers have $t_\mathrm{df}>1 \, \si{\giga\year}$, if the simulation was run to $z=0$ the number of numerical and delayed mergers would drastically increase.

In summary, BH mergers occur generally in galaxies with mass $M_\ast>10^9 \,\Msun$, avoiding therefore the low-mass end of the galaxy population. In a mass-matched galaxy sample, BH mergers have similar masses and accretion rates (in Eddington units) as the non-merging BHs, but they have higher mass and Eddington ratio than the full population including also BHs in low-mass galaxies. 

\subsubsection{Galaxy star formation rates}
\label{subsubsec:merger_sSFR}
\begin{figure}
	\includegraphics[width=\columnwidth]{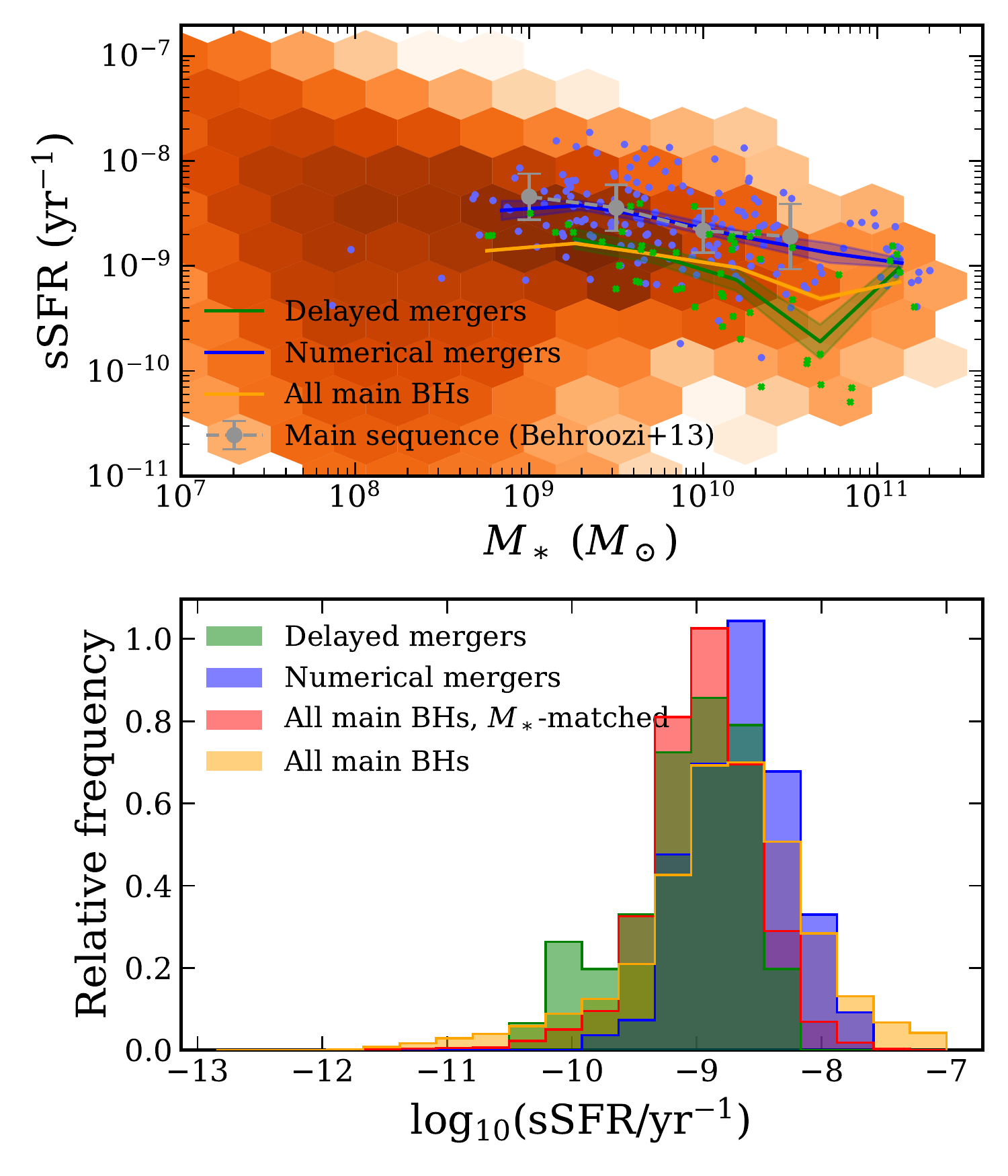}
    \caption{Star formation rate of BH merger hosts compared to the global population of galaxies. \textit{Top panel:} $\mathrm{sSFR}$ against stellar mass $M_\ast$ for delayed merger hosts (green crosses), numerical merger hosts (blue points), and the global population of main BH hosts (orange hexagons), at redshift $3.5<z<4$. For the hexagons, darker colours correspond to higher density of points. The geometric mean of $\mathrm{sSFR}$ in several bins of $M_\ast$ is shown by the solid lines for the three samples of BHs, with shaded regions corresponding to the $1\sigma$ standard error in the mean.
    \textit{Bottom panel:} The distribution of $\mathrm{sSFR}$ for delayed merger hosts, in green, numerical merger hosts, in blue, and for the global main 
    BH host population, in orange, at redshift $z = 3.5-4$. We show in red the global main BH host distribution weighted by a factor $\mathrm{pdf}_\mathrm{NM}(\log_{10}M_\ast)/\mathrm{pdf}_\mathrm{all}(\log_{10}M_\ast)$. Numerical mergers reside in galaxies with elevated $\mathrm{sSFR}$, a residue of SFR enhancement caused by the galaxy merger. Delayed mergers reside instead in galaxies compatible with the global population since sufficient time has elapsed from the galaxy merger for its effects to have vanished. }
    \label{fig:sSFR_mergers}
\end{figure}

The distribution of specific star formation rates ($\mathrm{sSFR}$) is shown in Fig.~\ref{fig:sSFR_mergers}. For the global BH host galaxy population, the $\mathrm{sSFR}$s are distributed along a wide range of values at low masses, while at larger masses ($M_\ast \gtrsim 10^9\, M_\odot$) they have settled on the star-forming main sequence with significantly lower scatter ($\sim 0.4\,\mathrm{dex}$). Since BH mergers tend to reside in $M_\ast \gtrsim 10^9\, M_\odot$ galaxies, BH merger hosts live in this thinner region in $\mathrm{sSFR}$.

At fixed $M_\ast$, the average $\mathrm{sSFR}$ of numerical merger hosts is consistently a factor of $\sim 2$ higher than the global population of BH hosts, in line with previous findings \citep{DeGraf2021}. That is, the distribution of $\mathrm{sSFR}$ for numerical mergers is higher than for a global $M_\ast$-matched sample. The galaxy merger that can generate a BH merger also causes tidal interactions and shocks in the galactic gas that lead to the compression of gas and a subsequent starburst \citep[e.g.][]{Hernquist1989}. 

We found that the $\mathrm{sSFR}$ of BH merger hosts tend to return to average values ${\sim 100}\,\si{\mega\year}$ after the galaxy merger. This average does not vary significantly with redshift or galaxy mass, and it is comparable to the time between galaxy merger and BH numerical merger (${\sim 70}\,\si{\mega\year}$ at $3.5<z<4$).
Numerical BH mergers, therefore are biased tracers of the galaxy SFR since they occur when the interaction-enhanced SFR is still active. 
The merger-induced SFR burst vanishes on a shorter timescale than the post-processed dynamical delays for delayed mergers, which have a median value of ${\sim 330}\,\si{\mega\year}$ at $3.5<z<4$. Consequently, at the later time of the delayed merger, the $\mathrm{sSFR}$ burst to has in general totally decayed. Indeed, delayed mergers do not show any significant excess in star formation with respect to the global population in Fig.~\ref{fig:sSFR_mergers}. That is, BHs mergers do not reside in galaxies with statistically boosted $\mathrm{sSFR}$ if dynamical delays are taken into account.

It must be noted that, as shown in Fig.~\ref{fig:sSFR_mergers}, the global galaxy population lies on average a factor of $\sim 2$ below the observational cosmic main sequence \citep{Behroozi2013}, outside the $1\sigma$ region. This is in contrast with the \textsc{NewHorizon} simulation \citep{Dubois2021}, where galaxies were found to be in good agreement with the observed main sequence. The physical modelling of \textsc{NewHorizon} closely resembles that of the \textsc{Obelisk}. Nevertheless, \textsc{NewHorizon} follows the evolution of an average patch of the Universe, while \textsc{Obelisk} evolves an overdense region. Quenching and feedback are more efficient in high-density regions, which likely causes the observed dearth of highly star-forming galaxies. Furthermore, for a direct comparison with observations, we would need to apply some luminosity cuts -- very low $\mathrm{sSFR}$ galaxies might not be picked up at high-$z$ in a UV-limited survey, therefore the observed average $\mathrm{sSFR}$ might be higher than for simulations if this cut is not applied in the simulated sample.

\subsubsection{BH spins}
\label{subsubsec:merger_spins}
\begin{figure}
	\includegraphics[width=\columnwidth]{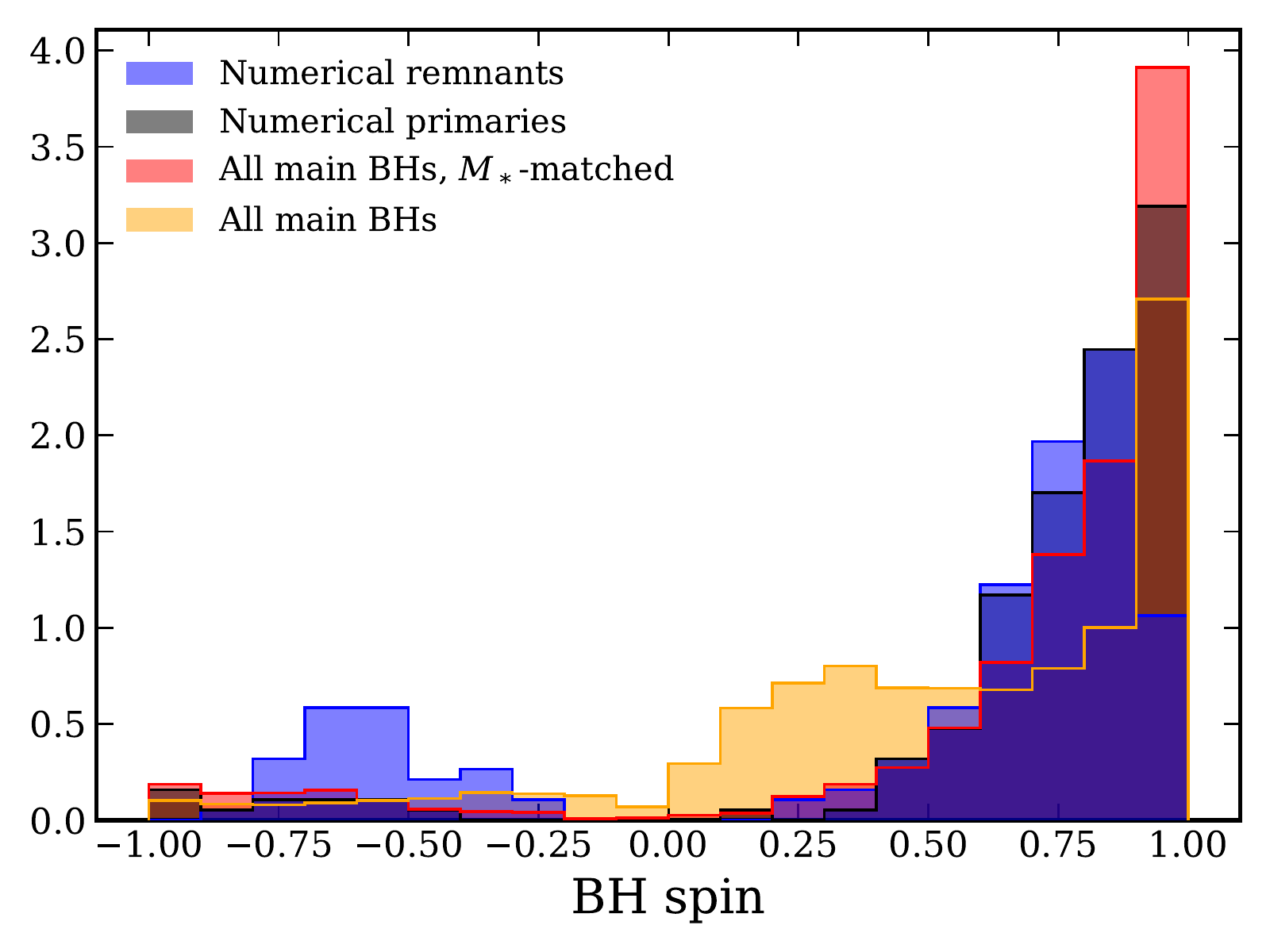}
    \caption{Distribution of BH spin for primary BHs in grey, for remnant BHs in blue, for the global main 
    BH population in orange, and for the global population weighted by the $M_\ast$ distribution of BH merger hosts, in red. All distributions correspond to redshift $3.5<z<4$. Pre-merger primaries have higher spins with respect to the global BH population but similar spins to an $M_\ast$-matched sample. Post-merger BHs have typically lower spin than prior to the merger, signifying that there is no full alignment. Post-merger BHs with negative spins generally result from spin flips at the merger. 
    }
    \label{fig:BHspin_mergers}

\end{figure}

As noted, we only consider numerical mergers for the spin analysis: this is because the spin of delayed mergers is modulated by further accretion and mergers during the dynamical delays. In Fig.~\ref{fig:BHspin_mergers} we explore the BH spin of merging BHs and the global population of main BHs. For numerical mergers, primary BHs generally have higher spin values compared to the global population of main BHs. Again, to a first approximation, the higher spins result from the different underlying $M_\ast$ distribution of their hosts -- the primary spin distribution is instead comparable with the mass-matched sample. As it was shown in Fig.~\ref{fig:merging_BH_distr}, merging BHs tend to have high $f_\mathrm{Edd}$ and gas accretion onto these BHs also tends to be coherent since the high-mass host galaxies are generally sufficiently mature to have formed coherently rotating structures (proto-discs or discs) in which gas is not constantly stirred by SN-induced turbulence \citep[see][and Sect.~\ref{sec:cosmic_evol_BHs}]{Dubois2015}. This increases the BH spin. 
Nonetheless, the spin distribution for primary BHs is slightly below that of the mass-matched galaxy distribution since BH mergers tend to be clustered in time, and thus merger primaries are likely to have their spins already decreased by recent mergers.

At the time of numerical coalescence, the BH spin of the remnant $a_f$ is determined by conservation of angular momentum -- the final spin is determined by summing over the contributions from the primary spin $a_1$, secondary spin $a_2$, and the orbital angular momentum $\ell$ \citep[][also see Eq.~\ref{eq:merger_spin}]{Rezzolla2008a}. The distribution of the resulting remnant spin distribution is shown in Fig.~\ref{fig:BHspin_mergers}. 

In comparison with the primary spin distribution, the post-merger spins are lower on average, as expected when there is no full alignment or a preference for strictly equatorial inspirals \citep{Berti2008}. In particular, there is a significant dearth of highly spinning BHs with $|a_f|>0.9$ and a larger population of BHs with a negative spin for the post-merger BH population. 
For primary BHs initially with high spins values, maintaining a high final spin requires a strong alignment of all other terms contributing significantly to the final spin.
A merger-induced change in the spin direction can also result in the misalignment with the gas angular momentum, which explains the higher number of negative spins. 

The contribution to the final spin from the orbital angular momentum and the secondary spin decrease with decreasing mass ratio $q$, while the contribution from the primary does not depend on $q$ (see Eq.~\ref{eq:merger_spin}). The modulus of the dimensionless orbital angular momentum term $\boldsymbol{\ell}$ is generally in the range \mbox{$2$ -- $4$}. We note that this is higher than the BH spins, which are normalised such that $a<1$. 
Consequently, the final spin has two regimes depending on the mass ratio. For $q \gtrsim 0.5$, the orbital angular momentum dominates the equation and dictates the direction of the final spin, with the BH spins contributing to a lesser degree. For smaller $q \lesssim 0.5$, the primary spin dominates, while it is perturbed mostly by the orbital term. It is important to note that the direction of the orbital angular momentum term $\boldsymbol{\ell}$ is not distributed uniformly, but tends to be aligned with the gas angular momentum, as expected for binaries in the presence of gas \citep{2007ApJ...661L.147B,Dotti2010}.

\begin{figure}
    \centering
    \includegraphics[width=\columnwidth]{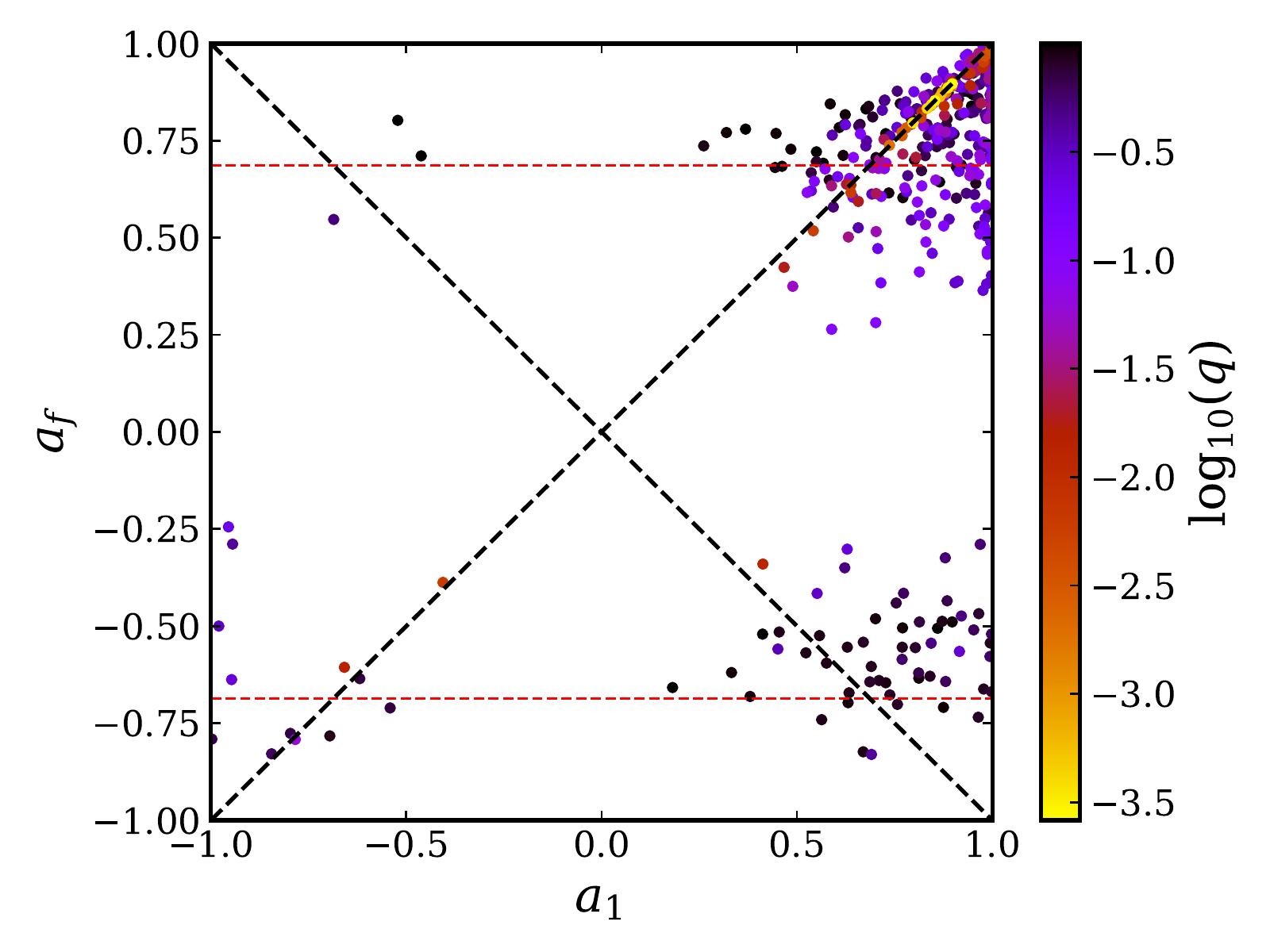}
    \caption{Spin of the numerical merger remnant against the primary spin prior to the merger, for mergers in the redshift range $3.5<z<4.5$. Each point indicates a merger, the colour encoding its mass ratio. Black dashed lines indicate the limit where the spin magnitude is unchanged, $a_f = a_1$ and $a_f=-a_1$. Red dashed lines at $a_f \sim \pm 0.69$ mark the final spin of an equal-mass merger of two non-spinning BHs. Two distinct populations can be observed in the diagram: high mass ratio ($q\gtrsim0.5$) mergers, mostly composed of low-mass BHs, and low mass ratio ($q\gtrsim0.5$) mergers, mostly composed of high-mass BHs.}
    \label{fig:af_vs_a1}
\end{figure}

In Fig.~\ref{fig:af_vs_a1}, we show the remnant BH spin in the time step immediately after the merger as a function of the primary BH spin before the merger. The two regimes described above create distinct populations that can be distinguished.

The $q \gtrsim 0.5$ sample is mostly composed of mergers of two low-mass BHs close to the simulation seed mass. This case accounts for $\sim 35\%$ of all mergers. We note that for the low-mass mergers which dominate this sample, host galaxies have a smaller gas content \citep{Barausse2012} and the gas dynamics are more erratic and dominated by SN-driven turbulence \citep{Dubois2015}, and so we expect little alignment between the angular momentum $\boldsymbol{\ell}$ and the initial BH spins. Consequently, post-merger spins are approximately distributed around $a_f \sim 0.69$ -- corresponding to the limiting case of an equal mass merger with $a_1=a_2=0$. The BH spins $\boldsymbol{a}_1$ and $\boldsymbol{a}_2$ add scatter around this mean value since the combination of the orbital and BH angular momenta can be constructive or destructive depending on their relative orientation. As the primary spin increases, the scatter of $a_f$ around the mean value also increases.
As observed in the figure, high $q$ mergers tend to increase the pre-merger spin for $a_1 \lesssim 0.69$ and to decrease it for $a_1 \gtrsim 0.69$.

In the $q \lesssim 0.5$ configuration, the main contributor to $a_f$ is the primary spin. Given the fixed BH seed mass in the simulation, low $q$ events correspond to primary BHs that have already grown and acquired higher spin values through gas accretion or mergers. The pre-merger primary spin is mostly perturbed by the orbital angular momentum. For the more massive host galaxies dominating this sample, gas dynamics tend to be more coherent. The orbital angular momentum $\boldsymbol{\ell}$ tends to be aligned with the gas angular momentum and the BH spins $\boldsymbol{a}_1$ and $\boldsymbol{a}_2$. Nonetheless, high alignment is required in order for the merger not to spin down the highly spinning BHs expected from such a coherent accretion model. Thus, the contribution from the orbital angular momentum tends to spin down the remnant BH for $a_1\gtrsim0.5$, which encompasses most of the merger events. It can also be noted that in the small $q\lesssim0.01$ limit, the merger becomes only a small perturbation and the remnant spin stays constant on the line $\boldsymbol{a}_f=\boldsymbol{a}_1$.

A significant fraction ($\sim 20\%$) of our remnant BHs experience a spin flip with respect to their progenitor primary BHs. Recall that spins are positive (negative) if the spin direction is aligned (anti-aligned) with the gas angular momentum. The mergers experiencing a spin flip are located in the top left (negative to positive) and bottom right (positive to negative) corner of Fig~\ref{fig:af_vs_a1}. We find predominantly positive-to-negative flips since there is a larger number of BHs with spins aligning with the accreted gas prior to the merger. Spin flips are caused mainly by a merger-induced change in the spin direction
and not by a change in the direction of the nuclear gas angular momentum. This occurs more frequently in the $q \gtrsim 0.5$ sample, where $\sim 35\%$ of mergers lead to a spin flip, 
compared to $\sim 5\%$ for $q \lesssim 0.5$. In the former case, spin flips are common since the final spin is dominated by $\boldsymbol{\ell}$, which is on average less aligned with the BH spins for the low-mass mergers dominating the $q \gtrsim 0.5$ sample. 
Spin flips can also be caused by a change in the direction of the gas angular momentum. However, we find this case to be subdominant in frequency since the direction of the accreted gas angular momentum generally remains coherent over the course of a coarse timestep. Spin-induced merger flips will reduce the radiative efficiency of the post-merger BHs, thus sharply changing their luminosity and will correlate with a high probability of BH ejection due to gravitational recoil \citep[e.g.][]{Campanelli2007,Dunn2020}.

We note that, if the merger-induced spin change were to take place at the delayed merger instead, as would be physically expected, the population of post-merger spins could be different than for numerical mergers. As found in Sect.~\ref{subsubsec:Mstar_MBH_and_fEdd}, delayed mergers tend to favour $q$ close to unity. Furthermore, as discussed in Sect.~\ref{subsec:merger_selection}, the evolution of the mass ratio during the binary phase is not well known theoretically, although recent numerical simulations of circumbinary accretion discs seem to suggest that the secondary can experience a larger accretion rate \citep{Farris2014,Duffell2020,Munoz2020} and drive the mass ratio towards unity. A higher fraction of mergers in the high $q$ regime, which is dominated by the binary angular momentum, would lead to a larger number of BHs with spin $\sim 0.69$ and a larger number of spin flips. Counteracting this effect, binaries embedded in circumbinary discs tend to have their spin and orbital angular momentum aligned with the angular momentum of the disc due to small-scale gas torques and the gas dynamical friction \citep{Berti2008,Dotti2010,Barausse2012}, which would lead to higher remnant spins.

\subsection{Redshift evolution of BH merger properties}
\label{subsec:redshift_evolution}

\begin{figure*}
    \includegraphics[width=\textwidth]{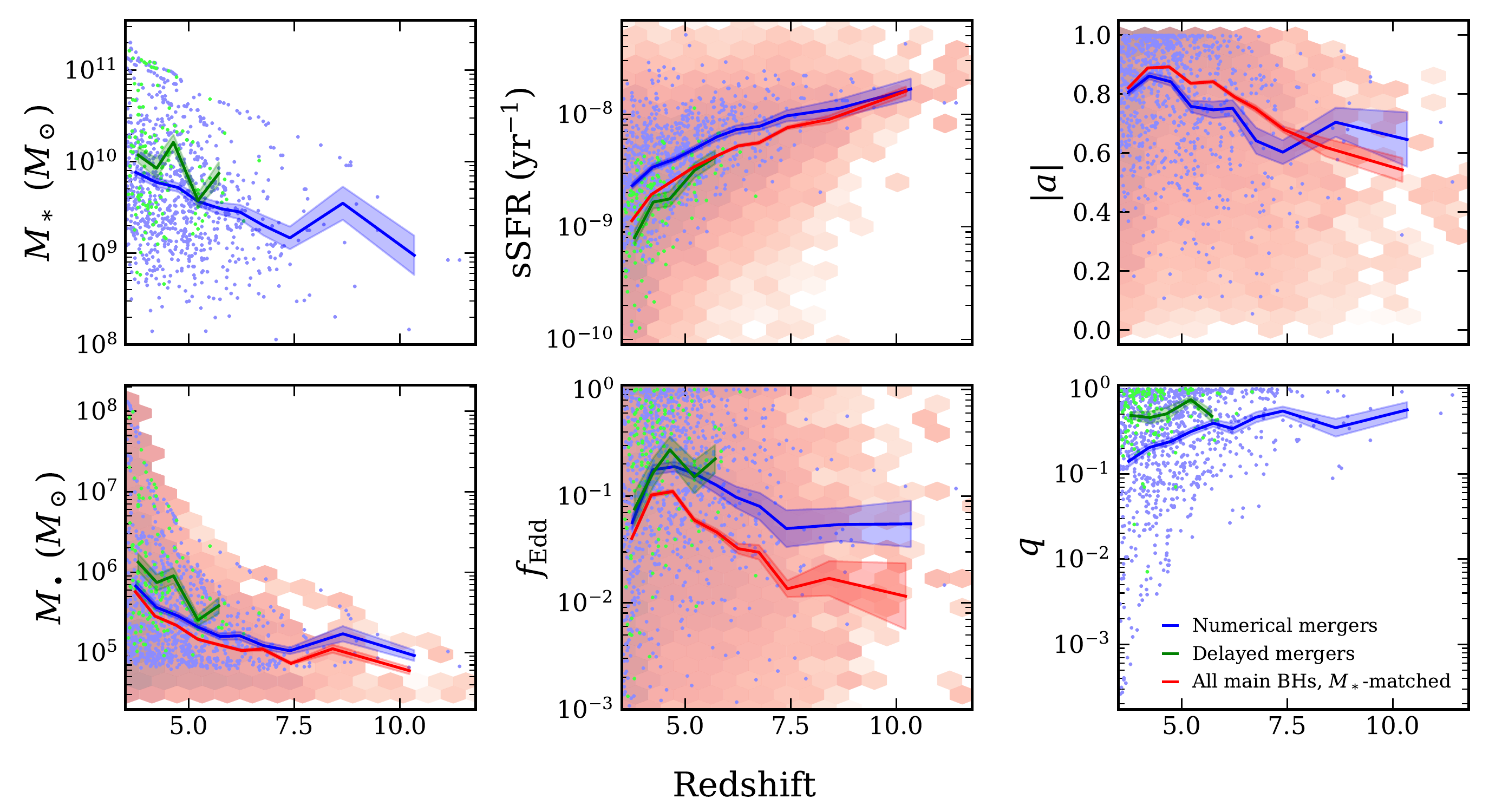}
    \caption{Distribution of host stellar mass, BH mass, specific star formation rate, BH Eddington ratio, BH spin, and BH mass ratio as a function of redshift for numerical mergers (blue points), delayed mergers (green points), and the global population of main BHs, matched to the $M_\ast$ distribution of numerical mergers by a factor $\mathrm{pdf}_\mathrm{NM}(\log_{10}M_\ast)/\mathrm{pdf}_\mathrm{all}(\log_{10}M_\ast)$ (red log-scaled hexagonal bins). For numerical mergers, we show the spin of the primary in the snapshot prior to the merger. For delayed mergers, the mass ratio refers to the pre-numerical merger mass ratio since the final mass ratio cannot be obtained from the simulation. The main BH population is matched to the $M_\ast$ distribution of numerical mergers for each redshift bin. In each bin, the geometric average of numerical mergers, delayed mergers, and mass-weighted main BHs is represented with solid lines, with their standard errors shown in the shaded regions. 
    }
    \label{fig:z_evol_mergers}
\end{figure*}
The properties of main BHs, merging BHs, and their hosts can vary significantly as a function of cosmic time, as shown in Fig.~\ref{fig:z_evol_mergers}.
As time progresses, BHs and galaxies build up their mass via mergers and gas consumption. Consequently, the average BH and galaxy mass increases as a function of time for numerical mergers (in blue), their $M_\ast$-weighted global population (in red), as well as for delayed mergers (in green). 
We note that here we $M_\ast$-match the global distribution by the $M_\ast$ distribution of numerical mergers since delayed mergers only appear at low redshift. We note however that the two $M_\ast$ distributions are qualitatively similar and bear similar results.

As gas forms stars, feeds BHs, and it is in turn affected by feedback from both, the amount of gas available for star formation decreases. This leads to a monotonic decrease in $\mathrm{sSFR}$ as a function of cosmic time. As in Fig~\ref{fig:sSFR_mergers}, the parent galaxy mergers tend to induce a $\mathrm{sSFR}$ enhancement for numerical mergers, except at the highest redshifts where mergers only weakly enhance star formation \citep{2017MNRAS.465.1934F}. 
The SFR boost has disappeared at the time of the delayed merger -- in fact, for delayed mergers, some galaxies are affected by a post-merger decline in $\mathrm{sSFR}$.

In the case of $f_\mathrm{Edd}$, we observe instead two competing trends -- at early times the build-up of BH and galaxy mass increases the average accretion efficiency, while at later times gas depletion is accelerated, driving a decrease in $f_\mathrm{Edd}$. Numerical mergers are found to accrete at a higher rate compared to a global $M_\ast$-matched sample, analogously to the $\mathrm{sSFR}$ boost. The galaxy merger can compress the nuclear gas and induce a $f_\mathrm{Edd}$ boost, especially for compact, high-$z$ galaxies. The amplitude of this boost decreases at lower redshifts. Even if dynamical delays are taken into account, mergers are still affected by a $f_\mathrm{Edd}$ boost. The initial galaxy merger-induced $f_\mathrm{Edd}$ boost allows BHs to grow to larger masses. Some BHs become overmassive and therefore accrete faster in the BHL model. As a consequence, delayed merger remnants have higher $f_\mathrm{Edd}$ at high $z$ compared to a $M_\ast$-matched sample (see Sect.~\ref{subsubsec:Mstar_MBH_and_fEdd}). Again, this boost becomes much smaller at lower redshift. 

Since the evolution of spin in our BH sample is mainly driven by gas accretion,
the average BH spin mimics the trend in $f_\mathrm{Edd}$, first increasing and then decreasing at later times.
The distribution of numerical merger primaries lies below the $M_\ast$-matched population since BH mergers tend to be clustered in time, and so previous recent mergers tend to have decreased the primary spin (as mentioned in Sect.~\ref{eq:merger_spin}).
The conclusions presented in Sect.~\ref{subsubsec:merger_spins} are found to apply in general to all redshifts. The primaries of BH mergers tend to have higher spins than the global population if no $M_\ast$-weighting is applied, and BH mergers tend to decrease the BH spin. The fraction of spin flips is also similar if all redshifts are taken into account, compared to the results presented above for low redshifts only.

As BHs increase their mass with time, the average mass ratio of numerical mergers decreases. The initial (pre-numerical merger) mass ratio for delayed mergers remains instead roughly constant through cosmic time since equal-mass mergers have shorter dynamical evolution timescales (see Sect.~\ref{subsubsec:Mstar_MBH_and_fEdd}). That is, BH systems with high mass ratios are in general not able to evolve fast enough to form a binary and coalesce. We note again that the final mass ratio of delayed mergers is difficult to model in post-processing.

\begin{figure}
	\includegraphics[width=\columnwidth]{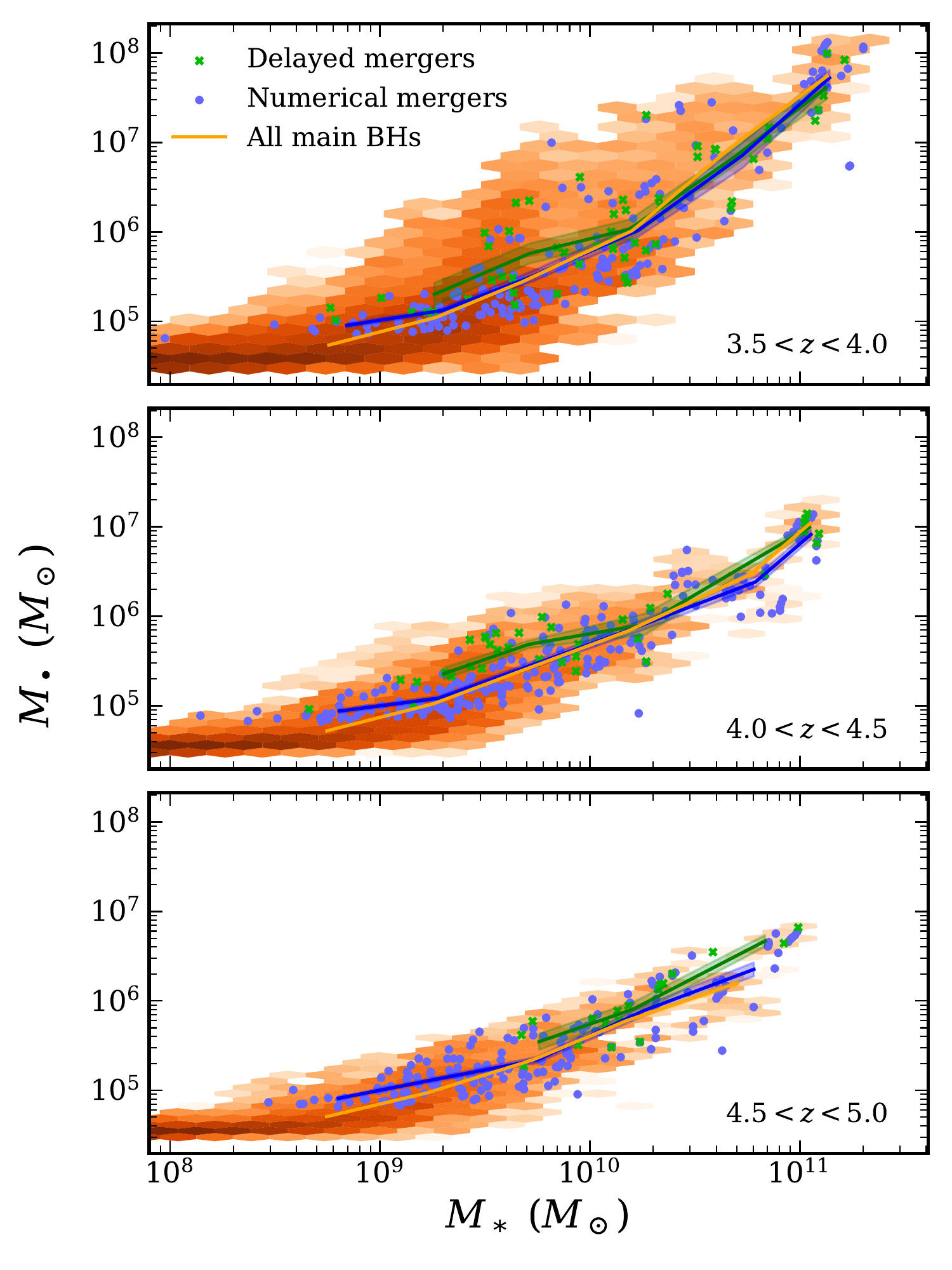}
    \caption{Correlation between BH mass and galaxy mass at different redshift ranges: $3.5<z<4$, $4<z<4.5$, and $4.5<z<5$. The notation is identical to  Fig.~\ref{fig:MBH_vs_Mgal_delay}. At constant $M_\ast$, $M_\bullet$ increases with redshift for $z<$. Delayed mergers are overmassive on average at certain galaxy masses at all redshifts, although the average lies within the scatter of the global relation.
    }
    \label{fig:MBH_vs_Mgal_zevol}
\end{figure}
As shown in Fig.~\ref{fig:MBH_vs_Mgal_zevol}, the correlation between $M_\bullet$ and $M_\ast$ also evolves with time. As discussed above and shown in Fig.~\ref{fig:z_evol_mergers}, the cosmic $\mathrm{sSFR}$ decreases monotonically with redshift, while the average $f_\mathrm{Edd}$ increases with time, reaches a peak at $z<5$, then decreases. Consequently, at early times galaxy growth dominates over BH growth, while at later times BH growth dominates. This late-time growth of BHs is show in Fig.~\ref{fig:MBH_vs_Mgal_zevol}, as the average $M_\bullet$ gets larger with cosmic time at constant $M_\ast$ for $M_\ast>10^{10} \,\Msun$. 
Delayed mergers are slightly overmassive with respect to the global relation at all redshifts considered here for some $M_\ast$ values, although they are within the scatter of the relationship when considering $\sigma$ instead of the error in the mean.

\section{Conclusions}
\label{sec:conclusions}

In this work, we have presented a comprehensive study of BH mergers in \textsc{Obelisk}, a cosmological hydrodynamical simulation following the evolution of a protocluster down to redshift $z \sim 3.5$. In particular, we have analysed the difference between properties of the population of numerical BH mergers (at the resolution of the simulation), delayed BH mergers (including post-processed delays below resolution), and the global population of main BHs. Our goal is to assess the possible biases that one may incur if the full BH population is inferred from the merging population. We summarise our results below: 
 \begin{itemize}
    \item The population of BH mergers is shaped by the astrophysical processes that determine BH cosmic evolution, which in turn is strongly influenced by the evolution of their host galaxies. At early times, the BH resides in a low-mass galaxy with no well-defined centre and erratic dynamics, leading to inefficient mass and spin growth. Later, for $M_\ast\gtrsim10^9\,\solarmass$ the galaxy is able to form a more regular structure, leading to a phase of efficient mass and spin growth due to gas accretion. At higher galaxy masses ($M_\ast\gtrsim10^{11}\,\solarmass$), the availability of gas decreases, and BH-BH mergers have a stronger impact on mass growth and spin evolution.
    
    \item The population of merging BHs presents significant differences with respect to the global underlying population (Fig.~\ref{fig:merging_BH_distr}). Merging BHs tend to reside in relatively massive galaxies ($\gtrsim10^9\,\solarmass$) and, at $z<4$, BH mass and accretion rate distributions for merging BHs are higher than the global main population but comparable for a $M_\ast$-matched sample. At higher redshift, BH masses and accretion rates remain instead higher for mergers than for a $M_\ast$-matched sample.
    Numerical mergers follow the global $M_\bullet-M_\ast$ relation, while delayed mergers are slightly overmassive on average in galaxies with $M_\ast <10^{10} \Msun$, although they agree within the scatter of the global relation (Fig.~\ref{fig:MBH_vs_Mgal_delay}).
    
    \item Taking into account sub-grid dynamical delays, the host galaxies of merging BHs have similar star formation properties as the full sample (Fig.~\ref{fig:sSFR_mergers}). Neglecting such delays would instead lead to selecting host galaxies $\sim 2$ times as star-forming as the full sample at fixed galaxy mass (Fig.~\ref{fig:sSFR_mergers}) except at the highest redshifts. This is caused by the star formation enhancement caused by the galaxy mergers that precede BH mergers. Delayed mergers are not affected by this star-formation boost because the delays tend to be considerably longer ($\sim 300\,\si{\mega\year}$ at $3.5<z<4$) than the star-formation boosts ($\sim 100\,\si{\mega\year}$). 

    \item   Numerical merger primary BHs tend to have higher spins compared to the full BH sample (Fig.~\ref{fig:BHspin_mergers}), but comparable spin values for a galaxy-mass-matched sample. The population of post-merger spins can be divided qualitatively into major ($q\gtrsim0.5$) and minor ($q\lesssim0.5$) mergers (Fig.~\ref{fig:af_vs_a1}). In major mergers, the final spin is dominated by the orbital angular momentum. Final spins tend to be distributed around $|a|\sim0.69$ and BHs often experience a spin flip. In minor mergers, the primary BH spin dominates, and the misalignment between the primary spin and the orbital angular momentum tends to decrease the final spin magnitude. The number of spin flips in this case is smaller.
    
    \item As galaxies and BHs are assembled from mergers and the consumption of gas, galaxy mass, BH mass, and BH spin all increase with cosmic time for both BH mergers and the full main BH sample (Fig.~\ref{fig:z_evol_mergers}), while the specific star formation rate decreases, as eventually does also $f_\mathrm{Edd}$. We find that BH mergers at high redshift have an $f_\mathrm{Edd}$ boost, which becomes weaker at lower redshifts. 
    
    \item  The mass ratio of BH mergers is a difficult quantity to predict: if no sub-grid delays are accounted for very unequal mass ratios become more probable as BHs grow in mass with time. When accounting for sub-grid delays the initial mass ratio remains close to unity at all redshifts since high mass ratios lead to short delay times. The evolution of the mass ratio all the way to coalescence has however too many uncertainties for a calculation in post-processing.
\end{itemize}

Further analysis of the BH merger population in \textsc{Obelisk} can be found in our companion paper \citet{Dong-Paez2023b}. We study the multi-messenger observability of our merger sample in both gravitational waves and the electromagnetic spectrum (radio, UV, and X-rays). We assess the complementarity of gravitational waves and the associated electromagnetic signals, as well as the possibility of transient signals which may aid the identification of a merger event. 

\begin{acknowledgements}
MV, YD, SV and NW acknowledge funding from the French National Research Agency (grant ANR-21-CE31-0026, project MBH\_waves). MT acknowledges support from the NWO grant 0.16.VIDI.189.162 (ODIN'). AM acknowledges support from the postdoctoral fellowships of IN2P3 (CNRS). This project has received funding from the European Union’s Horizon 2020 research and innovation programme
under the Marie Skłodowska-Curie grant agreement No. 101066346 (MASSIVEBAYES). This work has received funding from the Centre National d’Etudes Spatiales.
This work has made use of the Horizon Cluster hosted by Institut d’Astrophysique de Paris; we thank Stéphane Rouberol for running smoothly this cluster for us. We acknowledge PRACE for awarding us access to Joliot Curie at GENCI@CEA, France, which was used to run most of the simulations presented in this work. Numerical computations were partly performed on the DANTE platform, APC, France. Additionally, this work was granted access to the HPC resources of CINES under allocations A0040406955 and A0040407637 made by GENCI.

\end{acknowledgements}


\bibliography{references}

\begin{thebibliography}{80}
\expandafter\ifx\csname natexlab\endcsname\relax\def\natexlab#1{#1}\fi

\bibitem[{{Amaro-Seoane} {et~al.}(2023){Amaro-Seoane}, {Andrews}, {Arca Sedda},
  {Askar}, {Baghi}, {Balasov}, {Bartos}, {Bavera}, {Bellovary}, {Berry},
  {Berti}, {Bianchi}, {Blecha}, {Blondin}, {Bogdanovi{\'c}}, {Boissier},
  {Bonetti}, {Bonoli}, {Bortolas}, {Breivik}, {Capelo}, {Caramete},
  {Cattorini}, {Charisi}, {Chaty}, {Chen}, {Chru{\'s}li{\'n}ska}, {Chua},
  {Church}, {Colpi}, {D'Orazio}, {Danielski}, {Davies}, {Dayal}, {De Rosa},
  {Derdzinski}, {Destounis}, {Dotti}, {Du{\r{A}}{\textsterling}an}, {Dvorkin},
  {Fabj}, {Foglizzo}, {Ford}, {Fouvry}, {Franchini}, {Fragos}, {Fryer},
  {Gaspari}, {Gerosa}, {Graziani}, {Groot}, {Habouzit}, {Haggard}, {Haiman},
  {Han}, {Istrate}, {Johansson}, {Khan}, {Kimpson}, {Kokkotas}, {Kong},
  {Korol}, {Kremer}, {Kupfer}, {Lamberts}, {Larson}, {Lau}, {Liu},
  {Lloyd-Ronning}, {Lodato}, {Lupi}, {Ma}, {Maccarone}, {Mandel}, {Mangiagli},
  {Mapelli}, {Mathis}, {Mayer}, {McGee}, {McKernan}, {Miller}, {Mota},
  {Mumpower}, {Nasim}, {Nelemans}, {Noble}, {Pacucci}, {Panessa},
  {Paschalidis}, {Pfister}, {Porquet}, {Quenby}, {Ricarte}, {R{\"o}pke},
  {Regan}, {Rosswog}, {Ruiter}, {Ruiz}, {Runnoe}, {Schneider}, {Schnittman},
  {Secunda}, {Sesana}, {Seto}, {Shao}, {Shapiro}, {Sopuerta}, {Stone},
  {Suvorov}, {Tamanini}, {Tamfal}, {Tauris}, {Temmink}, {Tomsick}, {Toonen},
  {Torres-Orjuela}, {Toscani}, {Tsokaros}, {Unal}, {V{\'a}zquez-Aceves},
  {Valiante}, {van Putten}, {van Roestel}, {Vignali}, {Volonteri}, {Wu},
  {Younsi}, {Yu}, {Zane}, {Zwick}, {Antonini}, {Baibhav}, {Barausse}, {Bonilla
  Rivera}, {Branchesi}, {Branduardi-Raymont}, {Burdge}, {Chakraborty},
  {Cuadra}, {Dage}, {Davis}, {de Mink}, {Decarli}, {Doneva}, {Escoffier},
  {Gandhi}, {Haardt}, {Lousto}, {Nissanke}, {Nordhaus}, {O'Shaughnessy},
  {Portegies Zwart}, {Pound}, {Schussler}, {Sergijenko}, {Spallicci},
  {Vernieri}, \& {Vigna-G{\'o}mez}}]{2023LRR....26....2A}
{Amaro-Seoane}, P., {Andrews}, J., {Arca Sedda}, M., {et~al.} 2023, Living
  Reviews in Relativity, 26, 2

\bibitem[{{Angl{\'e}s-Alc{\'a}zar} {et~al.}(2017){Angl{\'e}s-Alc{\'a}zar},
  {Faucher-Gigu{\`e}re}, {Quataert}, {Hopkins}, {Feldmann}, {Torrey}, {Wetzel},
  \& {Kere{\v{s}}}}]{2017MNRAS.472L.109A}
{Angl{\'e}s-Alc{\'a}zar}, D., {Faucher-Gigu{\`e}re}, C.-A., {Quataert}, E.,
  {et~al.} 2017, \mnras, 472, L109

\bibitem[{{Arun} {et~al.}(2022)}]{2022LRR....25....4A}
{Arun}, K.~G. {et~al.} 2022, Living Reviews in Relativity, 25, 4

\bibitem[{{Aubert} {et~al.}(2004){Aubert}, {Pichon}, \& {Colombi}}]{Aubert2004}
{Aubert}, D., {Pichon}, C., \& {Colombi}, S. 2004, \mnras, 352, 376

\bibitem[{{Auclair} {et~al.}(2022)}]{2022arXiv220405434A}
{Auclair}, P. {et~al.} 2022, arXiv e-prints, arXiv:2204.05434

\bibitem[{{Ba{\~n}ados} {et~al.}(2018){Ba{\~n}ados}, {Venemans},
  {Mazzucchelli}, {Farina}, {Walter}, {Wang}, {Decarli}, {Stern}, {Fan},
  {Davies}, {Hennawi}, {Simcoe}, {Turner}, {Rix}, {Yang}, {Kelson}, {Rudie}, \&
  {Winters}}]{Banados2018}
{Ba{\~n}ados}, E., {Venemans}, B.~P., {Mazzucchelli}, C., {et~al.} 2018, \nat,
  553, 473

\bibitem[{{Barausse}(2012)}]{Barausse2012}
{Barausse}, E. 2012, \mnras, 423, 2533

\bibitem[{{Bardeen}(1970)}]{Bardeen1970}
{Bardeen}, J.~M. 1970, \nat, 226, 64

\bibitem[{{Beckmann} {et~al.}(2022){Beckmann}, {Dubois}, {Volonteri},
  {Dong-P{\'a}ez}, {Trebitsch}, {Devriendt}, {Kaviraj}, {Kimm}, \&
  {Peirani}}]{Beckmann2022}
{Beckmann}, R.~S., {Dubois}, Y., {Volonteri}, M., {et~al.} 2022, arXiv
  e-prints, arXiv:2211.13301

\bibitem[{{Begelman} {et~al.}(1980){Begelman}, {Blandford}, \&
  {Rees}}]{1980Natur.287..307B}
{Begelman}, M.~C., {Blandford}, R.~D., \& {Rees}, M.~J. 1980, \nat, 287, 307

\bibitem[{{Behroozi} {et~al.}(2013){Behroozi}, {Wechsler}, \&
  {Conroy}}]{Behroozi2013}
{Behroozi}, P.~S., {Wechsler}, R.~H., \& {Conroy}, C. 2013, \apj, 770, 57

\bibitem[{{Bellovary} {et~al.}(2019){Bellovary}, {Cleary}, {Munshi}, {Tremmel},
  {Christensen}, {Brooks}, \& {Quinn}}]{Bellovary2019}
{Bellovary}, J.~M., {Cleary}, C.~E., {Munshi}, F., {et~al.} 2019, \mnras, 482,
  2913

\bibitem[{{Berti} \& {Volonteri}(2008)}]{Berti2008}
{Berti}, E. \& {Volonteri}, M. 2008, \apj, 684, 822

\bibitem[{{Bogdanovi{\'c}} {et~al.}(2007){Bogdanovi{\'c}}, {Reynolds}, \&
  {Miller}}]{2007ApJ...661L.147B}
{Bogdanovi{\'c}}, T., {Reynolds}, C.~S., \& {Miller}, M.~C. 2007, \apjl, 661,
  L147

\bibitem[{{Bonetti} {et~al.}(2019){Bonetti}, {Sesana}, {Haardt}, {Barausse}, \&
  {Colpi}}]{2019MNRAS.486.4044B}
{Bonetti}, M., {Sesana}, A., {Haardt}, F., {Barausse}, E., \& {Colpi}, M. 2019,
  \mnras, 486, 4044

\bibitem[{{Bower} {et~al.}(2017){Bower}, {Schaye}, {Frenk}, {Theuns},
  {Schaller}, {Crain}, \& {McAlpine}}]{2017MNRAS.465...32B}
{Bower}, R.~G., {Schaye}, J., {Frenk}, C.~S., {et~al.} 2017, \mnras, 465, 32

\bibitem[{{Bustamante} \& {Springel}(2019)}]{2019MNRAS.490.4133B}
{Bustamante}, S. \& {Springel}, V. 2019, \mnras, 490, 4133

\bibitem[{{Byrne} {et~al.}(2023){Byrne}, {Faucher-Gigu{\`e}re}, {Stern},
  {Angl{\'e}s-Alc{\'a}zar}, {Wellons}, {Gurvich}, \& {Hopkins}}]{byrneetal23}
{Byrne}, L., {Faucher-Gigu{\`e}re}, C.-A., {Stern}, J., {et~al.} 2023, \mnras,
  520, 722

\bibitem[{{Campanelli} {et~al.}(2007){Campanelli}, {Lousto}, {Zlochower}, \&
  {Merritt}}]{Campanelli2007}
{Campanelli}, M., {Lousto}, C., {Zlochower}, Y., \& {Merritt}, D. 2007, \apjl,
  659, L5

\bibitem[{{Capelo} {et~al.}(2015){Capelo}, {Volonteri}, {Dotti}, {Bellovary},
  {Mayer}, \& {Governato}}]{2015MNRAS.447.2123C}
{Capelo}, P.~R., {Volonteri}, M., {Dotti}, M., {et~al.} 2015, \mnras, 447, 2123

\bibitem[{{Chen} {et~al.}(2022){Chen}, {Ni}, {Tremmel}, {Di Matteo}, {Bird},
  {DeGraf}, \& {Feng}}]{2022MNRAS.510..531C}
{Chen}, N., {Ni}, Y., {Tremmel}, M., {et~al.} 2022, \mnras, 510, 531

\bibitem[{{DeGraf} {et~al.}(2021){DeGraf}, {Sijacki}, {Di Matteo},
  {Holley-Bockelmann}, {Snyder}, \& {Springel}}]{DeGraf2021}
{DeGraf}, C., {Sijacki}, D., {Di Matteo}, T., {et~al.} 2021, \mnras, 503, 3629

\bibitem[{{Dong-P{\'a}ez} {et~al.}(2023){Dong-P{\'a}ez}, {Volonteri},
  {Beckmann}, {Dubois}, {Mangiagli}, {Trebitsch}, {Vergani}, \&
  {Webb}}]{Dong-Paez2023b}
{Dong-P{\'a}ez}, C.~A., {Volonteri}, M., {Beckmann}, R.~S., {et~al.} 2023,
  arXiv e-prints, arXiv:2303.09569

\bibitem[{{Dotti} {et~al.}(2015){Dotti}, {Merloni}, \&
  {Montuori}}]{2015MNRAS.448.3603D}
{Dotti}, M., {Merloni}, A., \& {Montuori}, C. 2015, \mnras, 448, 3603

\bibitem[{{Dotti} {et~al.}(2010){Dotti}, {Volonteri}, {Perego}, {Colpi},
  {Ruszkowski}, \& {Haardt}}]{Dotti2010}
{Dotti}, M., {Volonteri}, M., {Perego}, A., {et~al.} 2010, \mnras, 402, 682

\bibitem[{{Dubois} {et~al.}(2021){Dubois}, {Beckmann}, {Bournaud}, {Choi},
  {Devriendt}, {Jackson}, {Kaviraj}, {Kimm}, {Kraljic}, {Laigle}, {Martin},
  {Park}, {Peirani}, {Pichon}, {Volonteri}, \& {Yi}}]{Dubois2021}
{Dubois}, Y., {Beckmann}, R., {Bournaud}, F., {et~al.} 2021, \aap, 651, A109

\bibitem[{{Dubois} {et~al.}(2013){Dubois}, {Pichon}, {Devriendt}, {Silk},
  {Haehnelt}, {Kimm}, \& {Slyz}}]{Dubois2013}
{Dubois}, Y., {Pichon}, C., {Devriendt}, J., {et~al.} 2013, \mnras, 428, 2885

\bibitem[{{Dubois} {et~al.}(2014{\natexlab{a}}){Dubois}, {Pichon}, {Welker},
  {Le Borgne}, {Devriendt}, {Laigle}, {Codis}, {Pogosyan}, {Arnouts},
  {Benabed}, {Bertin}, {Blaizot}, {Bouchet}, {Cardoso}, {Colombi}, {de
  Lapparent}, {Desjacques}, {Gavazzi}, {Kassin}, {Kimm}, {McCracken},
  {Milliard}, {Peirani}, {Prunet}, {Rouberol}, {Silk}, {Slyz}, {Sousbie},
  {Teyssier}, {Tresse}, {Treyer}, {Vibert}, \& {Volonteri}}]{Dubois2014c}
{Dubois}, Y., {Pichon}, C., {Welker}, C., {et~al.} 2014{\natexlab{a}}, \mnras,
  444, 1453

\bibitem[{{Dubois} {et~al.}(2014{\natexlab{b}}){Dubois}, {Volonteri}, \&
  {Silk}}]{Dubois2014a}
{Dubois}, Y., {Volonteri}, M., \& {Silk}, J. 2014{\natexlab{b}}, \mnras, 440,
  1590

\bibitem[{{Dubois} {et~al.}(2014{\natexlab{c}}){Dubois}, {Volonteri}, {Silk},
  {Devriendt}, \& {Slyz}}]{Dubois2014b}
{Dubois}, Y., {Volonteri}, M., {Silk}, J., {Devriendt}, J., \& {Slyz}, A.
  2014{\natexlab{c}}, \mnras, 440, 2333

\bibitem[{{Dubois} {et~al.}(2015){Dubois}, {Volonteri}, {Silk}, {Devriendt},
  {Slyz}, \& {Teyssier}}]{Dubois2015}
{Dubois}, Y., {Volonteri}, M., {Silk}, J., {et~al.} 2015, \mnras, 452, 1502

\bibitem[{{Duffell} {et~al.}(2020){Duffell}, {D'Orazio}, {Derdzinski},
  {Haiman}, {MacFadyen}, {Rosen}, \& {Zrake}}]{Duffell2020}
{Duffell}, P.~C., {D'Orazio}, D., {Derdzinski}, A., {et~al.} 2020, \apj, 901,
  25

\bibitem[{{Dunn} {et~al.}(2020){Dunn}, {Holley-Bockelmann}, \&
  {Bellovary}}]{Dunn2020}
{Dunn}, G., {Holley-Bockelmann}, K., \& {Bellovary}, J. 2020, \apj, 896, 72

\bibitem[{{Fan} {et~al.}(2001){Fan}, {Narayanan}, {Lupton}, {Strauss}, {Knapp},
  {Becker}, {White}, {Pentericci}, {Leggett}, {Haiman}, {Gunn}, {Ivezi{\'c}},
  {Schneider}, {Anderson}, {Brinkmann}, {Bahcall}, {Connolly}, {Csabai}, {Doi},
  {Fukugita}, {Geballe}, {Grebel}, {Harbeck}, {Hennessy}, {Lamb}, {Miknaitis},
  {Munn}, {Nichol}, {Okamura}, {Pier}, {Prada}, {Richards}, {Szalay}, \&
  {York}}]{2001AJ....122.2833F}
{Fan}, X., {Narayanan}, V.~K., {Lupton}, R.~H., {et~al.} 2001, \aj, 122, 2833

\bibitem[{{Farris} {et~al.}(2014){Farris}, {Duffell}, {MacFadyen}, \&
  {Haiman}}]{Farris2014}
{Farris}, B.~D., {Duffell}, P., {MacFadyen}, A.~I., \& {Haiman}, Z. 2014, \apj,
  783, 134

\bibitem[{{Fensch} {et~al.}(2017){Fensch}, {Renaud}, {Bournaud}, {Duc},
  {Agertz}, {Amram}, {Combes}, {Di Matteo}, {Elmegreen}, {Emsellem}, {Jog},
  {Perret}, {Struck}, \& {Teyssier}}]{2017MNRAS.465.1934F}
{Fensch}, J., {Renaud}, F., {Bournaud}, F., {et~al.} 2017, \mnras, 465, 1934

\bibitem[{{Gabor} {et~al.}(2016){Gabor}, {Capelo}, {Volonteri}, {Bournaud},
  {Bellovary}, {Governato}, \& {Quinn}}]{2016A&A...592A..62G}
{Gabor}, J.~M., {Capelo}, P.~R., {Volonteri}, M., {et~al.} 2016, \aap, 592, A62

\bibitem[{{Habouzit} {et~al.}(2017){Habouzit}, {Volonteri}, \&
  {Dubois}}]{habouzitetal17}
{Habouzit}, M., {Volonteri}, M., \& {Dubois}, Y. 2017, \mnras, 468, 3935

\bibitem[{{Hernquist}(1989)}]{Hernquist1989}
{Hernquist}, L. 1989, \nat, 340, 687

\bibitem[{{Izquierdo-Villalba} {et~al.}(2022){Izquierdo-Villalba}, {Sesana},
  {Bonoli}, \& {Colpi}}]{Izquierdo-Villalba2022}
{Izquierdo-Villalba}, D., {Sesana}, A., {Bonoli}, S., \& {Colpi}, M. 2022,
  \mnras, 509, 3488

\bibitem[{{Katz} {et~al.}(2020){Katz}, {Kelley}, {Dosopoulou}, {Berry},
  {Blecha}, \& {Larson}}]{2020MNRAS.491.2301K}
{Katz}, M.~L., {Kelley}, L.~Z., {Dosopoulou}, F., {et~al.} 2020, \mnras, 491,
  2301

\bibitem[{{Kimm} \& {Cen}(2014)}]{kimm&cen14}
{Kimm}, T. \& {Cen}, R. 2014, \apj, 788, 121

\bibitem[{{King} {et~al.}(2005){King}, {Lubow}, {Ogilvie}, \&
  {Pringle}}]{King05}
{King}, A.~R., {Lubow}, S.~H., {Ogilvie}, G.~I., \& {Pringle}, J.~E. 2005,
  \mnras, 363, 49

\bibitem[{{Komatsu} {et~al.}(2011){Komatsu}, {Smith}, {Dunkley}, {Bennett},
  {Gold}, {Hinshaw}, {Jarosik}, {Larson}, {Nolta}, {Page}, {Spergel},
  {Halpern}, {Hill}, {Kogut}, {Limon}, {Meyer}, {Odegard}, {Tucker}, {Weiland},
  {Wollack}, \& {Wright}}]{Komatsu2011}
{Komatsu}, E., {Smith}, K.~M., {Dunkley}, J., {et~al.} 2011, \apjs, 192, 18

\bibitem[{{Kroupa}(2001)}]{Kroupa2001}
{Kroupa}, P. 2001, \mnras, 322, 231

\bibitem[{{Lacey} \& {Cole}(1993)}]{1993MNRAS.262..627L}
{Lacey}, C. \& {Cole}, S. 1993, \mnras, 262, 627

\bibitem[{{Lapiner} {et~al.}(2021){Lapiner}, {Dekel}, \&
  {Dubois}}]{Lapiner2021}
{Lapiner}, S., {Dekel}, A., \& {Dubois}, Y. 2021, \mnras, 505, 172

\bibitem[{{Li} {et~al.}(2020{\natexlab{a}}){Li}, {Bogdanovi{\'c}}, \&
  {Ballantyne}}]{2020ApJ...896..113L}
{Li}, K., {Bogdanovi{\'c}}, T., \& {Ballantyne}, D.~R. 2020{\natexlab{a}},
  \apj, 896, 113

\bibitem[{{Li} {et~al.}(2020{\natexlab{b}}){Li}, {Bogdanovi{\'c}}, \&
  {Ballantyne}}]{2020ApJ...905..123L}
{Li}, K., {Bogdanovi{\'c}}, T., \& {Ballantyne}, D.~R. 2020{\natexlab{b}},
  \apj, 905, 123

\bibitem[{{Luo} {et~al.}(2016){Luo}, {Chen}, {Duan}, {Gong}, {Hu}, {Ji}, {Liu},
  {Mei}, {Milyukov}, {Sazhin}, {Shao}, {Toth}, {Tu}, {Wang}, {Wang}, {Yeh},
  {Zhan}, {Zhang}, {Zharov}, \& {Zhou}}]{2016CQGra..33c5010L}
{Luo}, J., {Chen}, L.-S., {Duan}, H.-Z., {et~al.} 2016, Classical and Quantum
  Gravity, 33, 035010

\bibitem[{{Martin} {et~al.}(2018){Martin}, {Kaviraj}, {Volonteri}, {Simmons},
  {Devriendt}, {Lintott}, {Smethurst}, {Dubois}, \&
  {Pichon}}]{2018MNRAS.476.2801M}
{Martin}, G., {Kaviraj}, S., {Volonteri}, M., {et~al.} 2018, \mnras, 476, 2801

\bibitem[{{Mayer}(2017)}]{2017JPhCS.840a2025M}
{Mayer}, L. 2017, in Journal of Physics Conference Series, Vol. 840, Journal of
  Physics Conference Series, 012025

\bibitem[{{McAlpine} {et~al.}(2020){McAlpine}, {Harrison}, {Rosario},
  {Alexander}, {Ellison}, {Johansson}, \& {Patton}}]{mcalpine20}
{McAlpine}, S., {Harrison}, C.~M., {Rosario}, D.~J., {et~al.} 2020, \mnras,
  494, 5713

\bibitem[{{McKinney} {et~al.}(2012){McKinney}, {Tchekhovskoy}, \&
  {Blandford}}]{McKinney2012}
{McKinney}, J.~C., {Tchekhovskoy}, A., \& {Blandford}, R.~D. 2012, \mnras, 423,
  3083

\bibitem[{{Mu{\~n}oz} {et~al.}(2020){Mu{\~n}oz}, {Lai}, {Kratter}, \&
  {Miranda}}]{Munoz2020}
{Mu{\~n}oz}, D.~J., {Lai}, D., {Kratter}, K., \& {Miranda}, R. 2020, \apj, 889,
  114

\bibitem[{{Ni} {et~al.}(2022){Ni}, {Di Matteo}, {Bird}, {Croft}, {Feng},
  {Chen}, {Tremmel}, {DeGraf}, \& {Li}}]{Ni2022}
{Ni}, Y., {Di Matteo}, T., {Bird}, S., {et~al.} 2022, \mnras, 513, 670

\bibitem[{{Pfister} {et~al.}(2019){Pfister}, {Volonteri}, {Dubois}, {Dotti}, \&
  {Colpi}}]{Pfister2019}
{Pfister}, H., {Volonteri}, M., {Dubois}, Y., {Dotti}, M., \& {Colpi}, M. 2019,
  \mnras, 486, 101

\bibitem[{{Rezzolla} {et~al.}(2008){Rezzolla}, {Barausse}, {Dorband},
  {Pollney}, {Reisswig}, {Seiler}, \& {Husa}}]{Rezzolla2008a}
{Rezzolla}, L., {Barausse}, E., {Dorband}, E.~N., {et~al.} 2008, \prd, 78,
  044002

\bibitem[{{Rosdahl} {et~al.}(2013){Rosdahl}, {Blaizot}, {Aubert}, {Stranex}, \&
  {Teyssier}}]{Rosdahl2013}
{Rosdahl}, J., {Blaizot}, J., {Aubert}, D., {Stranex}, T., \& {Teyssier}, R.
  2013, \mnras, 436, 2188

\bibitem[{{Rosdahl} \& {Teyssier}(2015)}]{Rosdahl2015}
{Rosdahl}, J. \& {Teyssier}, R. 2015, \mnras, 449, 4380

\bibitem[{{Ruan} {et~al.}(2020){Ruan}, {Guo}, {Cai}, \&
  {Zhang}}]{2020IJMPA..3550075R}
{Ruan}, W.-H., {Guo}, Z.-K., {Cai}, R.-G., \& {Zhang}, Y.-Z. 2020,
  International Journal of Modern Physics A, 35, 2050075

\bibitem[{{Salcido} {et~al.}(2016){Salcido}, {Bower}, {Theuns}, {McAlpine},
  {Schaller}, {Crain}, {Schaye}, \& {Regan}}]{2016MNRAS.463..870S}
{Salcido}, J., {Bower}, R.~G., {Theuns}, T., {et~al.} 2016, \mnras, 463, 870

\bibitem[{{Sayeb} {et~al.}(2021){Sayeb}, {Blecha}, {Kelley}, {Gerosa},
  {Kesden}, \& {Thomas}}]{2021MNRAS.501.2531S}
{Sayeb}, M., {Blecha}, L., {Kelley}, L.~Z., {et~al.} 2021, \mnras, 501, 2531

\bibitem[{{Sesana} {et~al.}(2004){Sesana}, {Haardt}, {Madau}, \&
  {Volonteri}}]{2004ApJ...611..623S}
{Sesana}, A., {Haardt}, F., {Madau}, P., \& {Volonteri}, M. 2004, \apj, 611,
  623

\bibitem[{{Sesana} {et~al.}(2005){Sesana}, {Haardt}, {Madau}, \&
  {Volonteri}}]{2005ApJ...623...23S}
{Sesana}, A., {Haardt}, F., {Madau}, P., \& {Volonteri}, M. 2005, \apj, 623, 23

\bibitem[{{Sesana} \& {Khan}(2015)}]{2015MNRAS.454L..66S}
{Sesana}, A. \& {Khan}, F.~M. 2015, \mnras, 454, L66

\bibitem[{{Sesana} {et~al.}(2007){Sesana}, {Volonteri}, \&
  {Haardt}}]{2007MNRAS.377.1711S}
{Sesana}, A., {Volonteri}, M., \& {Haardt}, F. 2007, \mnras, 377, 1711

\bibitem[{{Shapiro}(2005)}]{2005ApJ...620...59S}
{Shapiro}, S.~L. 2005, \apj, 620, 59

\bibitem[{{Siwek} {et~al.}(2020){Siwek}, {Kelley}, \& {Hernquist}}]{Siwek2020}
{Siwek}, M.~S., {Kelley}, L.~Z., \& {Hernquist}, L. 2020, \mnras, 498, 537

\bibitem[{{Smethurst} {et~al.}(2022){Smethurst}, {Beckmann}, {Simmons}, {Coil},
  {Devriendt}, {Dubois}, {Garland}, {Lintott}, {Martin}, \&
  {Peirani}}]{Smethurst22}
{Smethurst}, R.~J., {Beckmann}, R.~S., {Simmons}, B.~D., {et~al.} 2022, arXiv
  e-prints, arXiv:2211.13677

\bibitem[{{Tamfal} {et~al.}(2018){Tamfal}, {Capelo}, {Kazantzidis}, {Mayer},
  {Potter}, {Stadel}, \& {Widrow}}]{Tamfal2018}
{Tamfal}, T., {Capelo}, P.~R., {Kazantzidis}, S., {et~al.} 2018, \apjl, 864,
  L19

\bibitem[{{Teyssier}(2002)}]{Teyssier2002}
{Teyssier}, R. 2002, \aap, 385, 337

\bibitem[{{Trebitsch} {et~al.}(2021){Trebitsch}, {Dubois}, {Volonteri},
  {Pfister}, {Cadiou}, {Katz}, {Rosdahl}, {Kimm}, {Pichon}, {Beckmann},
  {Devriendt}, \& {Slyz}}]{Trebitsch2021}
{Trebitsch}, M., {Dubois}, Y., {Volonteri}, M., {et~al.} 2021, \aap, 653, A154

\bibitem[{{Tremmel} {et~al.}(2015){Tremmel}, {Governato}, {Volonteri}, \&
  {Quinn}}]{Tremmel2015}
{Tremmel}, M., {Governato}, F., {Volonteri}, M., \& {Quinn}, T.~R. 2015,
  \mnras, 451, 1868

\bibitem[{{Tweed} {et~al.}(2009){Tweed}, {Devriendt}, {Blaizot}, {Colombi}, \&
  {Slyz}}]{Tweed2009}
{Tweed}, D., {Devriendt}, J., {Blaizot}, J., {Colombi}, S., \& {Slyz}, A. 2009,
  \aap, 506, 647

\bibitem[{{Volonteri} {et~al.}(2003){Volonteri}, {Haardt}, \&
  {Madau}}]{2003ApJ...582..559V}
{Volonteri}, M., {Haardt}, F., \& {Madau}, P. 2003, \apj, 582, 559

\bibitem[{{Volonteri} {et~al.}(2020){Volonteri}, {Pfister}, {Beckmann},
  {Dubois}, {Colpi}, {Conselice}, {Dotti}, {Martin}, {Jackson}, {Kraljic},
  {Pichon}, {Trebitsch}, {Yi}, {Devriendt}, \& {Peirani}}]{Volonteri2020}
{Volonteri}, M., {Pfister}, H., {Beckmann}, R.~S., {et~al.} 2020, \mnras, 498,
  2219

\bibitem[{{Wang} {et~al.}(2021){Wang}, {Yang}, {Fan}, {Hennawi}, {Barth},
  {Banados}, {Bian}, {Boutsia}, {Connor}, {Davies}, {Decarli}, {Eilers},
  {Farina}, {Green}, {Jiang}, {Li}, {Mazzucchelli}, {Nanni}, {Schindler},
  {Venemans}, {Walter}, {Wu}, \& {Yue}}]{Wang2021}
{Wang}, F., {Yang}, J., {Fan}, X., {et~al.} 2021, \apjl, 907, L1

\bibitem[{{Wang} {et~al.}(2019){Wang}, {Yang}, {Fan}, {Wu}, {Yue}, {Li},
  {Bian}, {Jiang}, {Ba{\~n}ados}, {Schindler}, {Findlay}, {Davies}, {Decarli},
  {Farina}, {Green}, {Hennawi}, {Huang}, {Mazzuccheli}, {McGreer}, {Venemans},
  {Walter}, {Dye}, {Lyke}, {Myers}, \& {Nunez}}]{Wang2019}
{Wang}, F., {Yang}, J., {Fan}, X., {et~al.} 2019, \apj, 884, 30

\bibitem[{{Yang} {et~al.}(2021){Yang}, {Wang}, {Fan}, {Barth}, {Hennawi},
  {Nanni}, {Bian}, {Davies}, {Farina}, {Schindler}, {Ba{\~n}ados}, {Decarli},
  {Eilers}, {Green}, {Guo}, {Jiang}, {Li}, {Venemans}, {Walter}, {Wu}, \&
  {Yue}}]{Yang2021}
{Yang}, J., {Wang}, F., {Fan}, X., {et~al.} 2021, \apj, 923, 262

\end{thebibliography}

\end{document}